\documentclass[aps,prb,amsmath,amssymb,superscriptaddress,showpacs,twocolumn,floatfix]{revtex4}
\usepackage{amsfonts, relsize}
\usepackage{graphics}
\usepackage{graphicx}
\usepackage{hyperref}

\usepackage{color}

\usepackage{hyperref}

\begin{document}

\title{$Z_2$ index for gapless fermionic modes in the vortex core of three dimensional paired Dirac fermions}

\author{Bitan Roy,}
\affiliation{ National High Magnetic Field Laboratory, Florida State University, FL 32310, USA}
\affiliation{ Condensed Matter Theory Center, Department of Physics, University of Maryland, College Park, MD 20742, USA}

\author{Pallab Goswami}
\affiliation{ National High Magnetic Field Laboratory, Florida State University, FL 32310, USA}
\date{\today}

\begin{abstract}
We consider the gapless modes along the vortex line of the fully gapped, momentum independent paired states of three-dimensional Dirac fermions. For this, we require the solution of fermion zero modes of the corresponding two-dimensional problem in the presence of a point vortex, in the plane perpendicular to the vortex line. Based on the spectral symmetry requirement for the existence of the zero mode, we identify the appropriate generalized Jackiw-Rossi Hamiltonians for different paired states. A four-dimensional generalized Jackiw-Rossi Hamiltonian possesses spectral symmetry with respect to an antiunitary operator, and gives rise to a single zero mode only for the {\em odd vorticity}, which is formally described by a $Z_2$ index. In the presence of generic perturbations such as chemical potential, Dirac mass, and Zeeman couplings, the associated two-dimensional problem for the odd parity topological superconducting state maps onto {\em two} copies of generalized Jackiw-Rossi Hamiltonian, and consequently an odd vortex binds two Majorana fermions. In contrast, there are no zero energy states for the topologically trivial $s$-wave superconductor in the presence of any chiral symmetry breaking perturbation in the particle-hole channel, such as regular Dirac mass. We show that the number of one-dimensional dispersive modes along the vortex line is also determined by the index of the associated two-dimensional problem. For an axial superfluid state in the presence of various perturbations, we discuss the consequences of the $Z_2$ index on the anomaly equations.
\end{abstract}


\maketitle

\section{introduction}

The existence of zero modes for the Dirac fermions in the presence of a topologically non-trivial configuration of an order parameter or a gauge field is an interesting problem in condensed matter and high energy physics. In a seminal paper, Jackiw and Rebbi demonstrated the emergence of the fermion zero modes for various defects in odd spatial dimensions \cite{Jackiw-Rebbi}. In particular, they have showed the existence of zero mode for a domain wall in one dimension, t'Hooft-Polyakov monopole and dyon in three dimensions. The fermion zero modes due to a solitonic defect can give rise to induced quantum numbers and fractionalization \cite{Jackiw-Rebbi, goldstone-wilczek}. The domain wall of a scalar order parameter (Dirac mass) in one dimension binds single zero mode, leading to the \emph{fractionalization of charge}. Zero-energy states bound to domain wall can be observed experimentally, for example, in polyacetylene \cite{Su-Schrieffer-Heeger}. Interestingly, the edge and the surface states of many gapped topological systems in higher dimensions are also determined by one or multiple copies of the one-dimensional Jackiw-Rebbi model. Some specific examples are the edge states of the $d+id$ \cite{Senthil} and the $p+ip$ \cite{ReadGreen, Stone-Roy} superconductors and the quantum spin Hall state \cite{qsh} in two dimensions, and the surface states of the three-dimensional topological insulators and superconductors \cite{TI-RMP}. For these higher-dimensional problems, the domain wall describes a boundary between topologically distinct vacua. It is also interesting to note that the chiral surface states of gapless Weyl semimetal and superconductors are also governed by similar one-dimensional problem \cite{tewari-goswami}. \\

The point vortex and the line vortex of a U(1) symmetry breaking order parameter respectively in two and three dimensions, are also interesting topological defects. The point vortices of the $p+ip$ superconductor \cite{ReadGreen, VolovikJetp, stone-chung} and the Kekule bond-density wave order in graphene \cite{hou-chamon-mudry} support localized fermion zero modes. In contrast, the line vortex of many three-dimensional paired states such as axial superfluid \cite{callan-harvey}, axial superconductor\cite{semenoff}, superfluid $^3$He-B \cite{volovik}, the gapped superconducting states of three-dimensional Dirac fermions \cite{nishida, Tagliacozzo, Yuki, Goswami-Roy-effectivetheory} support one-dimensional gapless fermions along the vortex core. If we consider, the line vortex along the $\hat{z}$ direction, the momentum $k_z$ is a conserved quantity, and the dispersing states are localized in the $x-y$ plane. Therefore, for $k_z=0$, we indeed encounter a two-dimensional problem of the fermion zero modes. Consequently, the number of dispersing modes becomes equal to the number of fermion zero modes of an effective two-dimensional problem in the presence of a point vortex.\\

The problem of zero-energy modes bound by a point vortex in two dimensions has been considered by Jackiw and Rossi \cite{JR-original, jackiw-pi}. The pertinent Hamiltonian is
\begin{equation}\label{JRintro}
H_{JR}\left[\mathbf{a}\right]= \sum_{j=1,2} \Gamma_j (-i\partial_j + i \Gamma_3 \Gamma_4 a_j) + \Delta_1 \Gamma_3 + \Delta_2 \Gamma_4 ,
\end{equation}
where $\partial_j$'s are the derivatives with respect to the spatial coordinates, and $\Delta_1, \Delta_2$ are the components of a complex order parameter. The four-dimensional Hermitian $\Gamma$-matrices satisfy the anticommuting algebra $\left\{ \Gamma_i, \Gamma_j \right\}= 2 \delta_{i j}$, where $i,j=1,2,3,4$. The gauge field corresponding to the broken $U(1)$ symmetry is denoted by $a_j$, and the finite amplitude of the order parameter gives rise to the Meissner effect for $a_j$. The point vortex of the order parameter is introduced as
\begin{equation}\label{vortdef}
\Delta_1+i \Delta_2 = |\Delta(r)| \; e^{i n \phi},
\end{equation}
where $n$ is the vorticity, and $\phi$ is the azimuthal angle, and $|\Delta(r)|$ describes the radial variation of the amplitude. It has been shown via a direct calculation \cite{JR-original} that there are exactly $n$ number of zero modes for vorticity $n$. Subsequently, the correspondence between the vorticity and the number of the zero energy states, has been expressed as a $Z$ index theorem by Weinberg \cite{weinberg-index}. This is an index defined on the open space \cite{Niemi-Semenoff} like Callias index theorem \cite{Callias}, unlike the celebrated Atiyah-Singer index theorem\cite{atiyah-singer}, which is defined on the compact manifold. There is an additional unitary matrix $\Gamma_5=\Gamma_1\Gamma_2\Gamma_3\Gamma_4$, which anti-commutes with $H_{JR}$, and ensures its spectral symmetry. All together the \emph{five} mutually anti-commuting $\Gamma$-matrices close the Clifford algebra of the four-dimensional matrices \cite{okubo}. As a consequence of the spectral symmetry, the zero energy states become eigenstates of $\Gamma_5$, with eigenvalues $\pm 1$. Therefore $\Gamma_5$ defines the \emph{chirality} of the zero modes. The $Z$ index theorem states that
\begin{equation}
N_+-N_-=n,
\end{equation}
where $N_{\pm}$ are respectively the number of zero energy states with chirality $\pm 1$. In the Altland-Zirnbauer classification scheme, this Hamiltonian ($H_{JR}$) belongs to the class BDI \cite{Altland, 10foldclassification}. The presence of the gauge field is not important for the number of zero modes or the associated index theorem. However, the existence of the gauge field is of paramount importance to realize deconfined vortices\cite{jackiw-pi}. \\

The Jackiw-Rossi Hamiltonian $H_{JR}\left[\mathbf{a}\right]$ can describe both insulating and superconducting states. For an insulator, $a_j$'s are the components of the chiral gauge field. In contrast, for a superconductor, $a_j$'s describe the physical electromagnetic vector potential. For the insulating states of the charged fermions, the electromagnetic U(1) symmetry is preserved and the $H_{JR}[\mathbf{a}]$ can be augmented by the electromagnetic vector potential $\mathbf{A}$, without altering the BDI chiral symmetry \cite{roy-realaxial-zeromode}. The modified Hamiltonian is
\begin{equation}
H_{JR} \left[ \mathbf{a},\mathbf{A} \right]= \sum_{j=1,2} \Gamma_j (-i\partial_j + i \Gamma_3 \Gamma_4 a_j + A_j) +  \Delta_1 \Gamma_3 + \Delta_2 \Gamma_4.
\end{equation}
The vector potential $\mathbf{A}$ only changes the wave functions of the zero modes, without affecting its number \cite{roy-realaxial-zeromode, herbut-realmagneticfield}. In particular for the insulators, when the vorticity is \emph{one} ($n=1$), there exists \emph{single} state at zero energy. This leads to the charge fractionalization, which is relevant for the spinless fermions on the honeycomb lattice with an underlying vortex of the Kekule bond density wave order \cite{hou-chamon-mudry, chamon-hou-semenoff-pi-jackiw, hou-chamon-mudry-magnetic-field, semenoff-realzeromodes}. The axial/chiral magnetic field (time-reversal preserving) can be introduced on the honeycomb lattice, for example, by deliberate wrinkling of the flake \cite{graphene-pseudo-magnetic-field}. For graphene, due to the spinful nature of the fermions, the charge fractionalization is lost. Instead, there is a competing antiferromagnetic order in the Kekule vortex core \cite{herbut-isospin}. The fermion zero modes also occur for the vortices of the easy plane antiferromagnetic and quantum spin Hall orders, which respectively support bond density wave order and $s$-wave pairing inside the vortex core \cite{roy-realaxial-zeromode, herbut-neel, herbut-realmagneticfield}.
\\

The fermion zero mode for the $s$-wave pairing of the topological insulator's surface states is also described by $ H_{JR}\left[\mathbf{a}\right]$ (by ignoring the chemical potential and the Zeeman coupling). Due to the BdG nature of the quasiparticles, the single zero energy mode corresponds to a \emph{Majorana fermion} \cite{fu-kane, ghaemi-aswin}. The superconductivity on the helical surface states of a topological insulator has been realized by the proximity effect \cite{Molenkamp}, and the possibility of realizing the Majorana zero mode is also being actively pursued \cite{goldhabar}. Similar considerations can be applied for the superconducting states on the honeycomb lattice \cite{ghaemi-wilczek, SCvortex-graphene-herbut, SCvortex-graphene-ghaemi, kekule-SC-graphene}. However, the minimal representation of spinful Dirac fermions in graphene is eight-dimensional and to accommodate the pairings, the Nambu's doubled representation becomes 16-dimensional. In this context, the associated Hamiltonian is a \emph{direct sum} of multiple copies of the \emph{four-dimensional} Jackiw-Rossi Hamiltonian, shown in Eq. (\ref{JRintro}). The existence of multiple zero energy modes even with $n=1$, ruins the possibility of realizing Majorana fermions \cite{chamon-majorana}. Nevertheless, the presence of the zero modes leads to an interesting interplay of various competing order parameters inside the vortex core \cite{herbut-isospin}.
\\

So far, we have been ignoring the effects of the finite chemical potential and the Zeeman coupling on the zero modes in a superconductor. Due to the inevitability of these couplings in a real system, it becomes important to ask which generalization of $H^{JR}[\mathbf{a}]$ still supports zero modes. At least for a single vortex (i.e., $n=1$), Herbut and Lu have provided the answer to this question \cite{herbut-lu-genJR}. It has been argued that one can introduce \emph{two} additional parameters to the original Jackiw-Rossi Hamiltonian in Eq. (\ref{JRintro}), which together can support zero energy mode, when $n=1$ in Eq.~(\ref{vortdef}). The generalized Jackiw-Rossi Hamiltonian then takes the form
\begin{equation}\label{genJR}
H^{JR}_{gen}= H_{JR} \left[ \mathbf{a} \right] + i \Gamma_3 \Gamma_4 \lambda + i \Gamma_1 \Gamma_2 \chi.
\end{equation}
It has been shown in Ref.~\onlinecite{herbut-lu-genJR} that there exists an antiunitary operator ($A$), which anticommutes with the total Hamiltonian $H^{JR}_{gen}$, guaranteeing its spectral symmetry. In the same work authors have showed that a point vortex of vorticity $\pm 1$ can bind one zero energy state only if $\Delta^2_0 + \lambda^2 > \chi^2$, where $\Delta_0 = |\Delta (r \to \infty)| $. In this work we search for the zero energy states in the spectrum of $H^{JR}_{gen}$ for arbitrary vorticity, and report an index theorem connecting the vorticity and number of zero modes. The physical meaning of the parameters ($\lambda, \chi$) are again representation dependent. For example, if we consider spinless fermions in monolayer graphene, $\lambda$ corresponds to a \emph{chiral chemical potential}, whereas $\chi$ to the Haldane anomalous mass \cite{Haldane}. On the other hand, for s-wave paring of the surface states of topological insulator, they respectively correspond to the ordinary chemical potential and the Zeeman coupling \cite{grosfeld-seradjeh-Vishveshwara}. It has been argued recently that the generalized Jackiw-Rossi Hamiltonian with $\chi=0$ can support a single zero energy mode only if the vorticity is \emph{odd}\cite{fukui-fujiwara}, and all the states are at finite energies for the even vorticity (except at $\lambda=\pm \chi$ \cite{santos-chamon}). A semiclassical treatement to this problem has also suggested the possibility of realizing zero modes only for odd vorticity \cite{teo-kane}. Since the spectral symmetry with respect to $\Gamma_5$ (unitary) is now lost, the concept of chirality and $Z$ index becomes moot. Therefore, the generalized Jackiw-Rossi Hamiltonian has a $Z_2$ index for the generic parameters ($\lambda \neq \pm \chi$), which can be stated as
\begin{equation}\label{indexstate}
\mathcal{N}=\frac{1}{2}\; \bigg(1-(-1)^n\bigg) \; \Theta \left ( \Delta^2_0 + \lambda^2 - \chi^2\right),
\end{equation}
where $\mathcal{N}$ and $n$ are, respectively, the number of zero modes and the vorticity. A similar $Z_2$ index also emerges for the two-dimensional $p+i p$ superconductors \cite{stone-chung, tewari-dassarma-lee, gurarie-radzihovsky}, which has been argued to be a nonrelativistic limit of the $H^{JR}_{gen}$ \cite{ santos-chamon, silavel-volovik}. The index for the zero modes in the spectrum of generalized Jackiw-Rossi Hamiltonian, $\mathcal{N}$, can also be presented as $\mathcal{N}=(1-C^{2D}_1)$, where $C^{2D}_1$ is the first Chern number in two spatial dimensions, given by\cite{santos-chamon}
\begin{equation}\label{2DChern}
C^{2D}_1=\left\{ 
\begin{array}{r l}
 0 \: \: \mbox{when} \: \: \Delta^2_0 + \lambda^2 > \chi^2, \\ \\
1 \: \:  \mbox{when} \: \:  \chi^2> \Delta^2_0 + \lambda^2.
\end{array} \right.
\end{equation}
Therefore, the zero mode index $\mathcal{N}$ can only be nonzero, when the first Chern number vanishes. \\

In this paper, we will mainly focus on the time reversal symmetric, fully gapped, momentum independent superconducting states of three-dimensional Dirac fermions, which are realized in many narrow gap and gapless semiconductors. In particular, there has been a surge of interest in the superconducting states of e.g. copper intercalated bismuth selenide $Cu_x Bi_2 Se_3$\cite{cubise}, indium doped tin telluride $Sn_{1-x} In_x Te$ \cite{insnte}. Apart from the regular s-wave pairing, the three-dimensional Dirac nature of the quasi-particle also allows the possibility of a fully gapped topologically nontrivial odd-parity pairing.\cite{liangfu} When the time reversal symmetry of the system is present the odd-parity pairing belongs to class DIII of Altland-Zirnbauer classification\cite{Altland, 10foldclassification}. Even though the pairing symmetry in these materials has not been established,\cite{maryland-exp} it is known that the paired state is fully gapped and of type-II nature.  This motivates us to study the possible dispersive modes along the vortex core in the mixed phase of these materials, and for concreteness we restrict ourselves in the dilute vortex limit (close to $H_{c1}$). \\

Now we provide a synopsis of our main findings in this paper.
\begin{enumerate}

\item For the $H^{JR}_{gen}$, we explicitly demonstrate the algebraic reason for the absence of the zero modes in the case of even vorticity. In addition, we reconfirm the existence of the single zero mode for odd vorticity.\\

\item We first consider the pairing of three-dimensional massless Dirac fermions in the absence of the chemical potential and  all other fermion bilinears in the particle-hole channel. When projected to the $x-y$ plane ($k_z=0$), the Hamiltonians for both s-wave and the topological odd parity pairings  map to two copies of $H_{JR}[0]$. Therefore the $Z$ index theorem of $H_{JR}[0]$ governs the number of Majorana zero modes.\\

\item When we incorporate the chemical potential, the Hamiltonian for both pairings map onto two copies of $H^{JR}_{gen}$. Consequently, a pair of Majorana zero modes are only found for odd vorticity. In addition, if we introduce the Dirac mass and the Zeeman couplings, only the odd parity topological pairing still leads to two copies of $H^{JR}_{gen}$. On the other hand, the effective Hamiltonian for the s-wave pairing falls outside the paradigm of $H^{JR}_{gen}$ in the presence of any chiral symmetyry breaking perturbations in the particle-hole channel (for example, scalar Dirac mass). Consequently, the zero modes are absent for $s$-wave pairing under generic perturbations. In contrast, the zero modes for the topological superconductors are more robust.\\

\item For the topological superconductor, $k_z$ causes mixing between the two Majorana modes and converts them into dispersing complex fermions. We provide a symmetry argument to explain why $k_z$ does not cause any mixing between the zero modes and the finite energy Caroli-de Gennes-Matricon states. Due to this reason, the degenerate perturbation theory for $k_z$ in the zero energy subspace provides an exact answer for the dispersion relation. Consequently, the $Z_2$ index associated with the number of zero modes bound to a point vortex in two-dimensions also dictates the index of the gapless modes along the line vortex. We also demonstrate this through some exact solutions.\\

\item In the concluding sections, we consider the generalization of the zero-mode problem in the context of \emph{axial superfluid}. We show that the fermion zero modes can be found in the presence of the axial chemical potential and the third component (along the vortex line) of the axial vector, for odd vorticity. This leads to an interesting modification of the Callan-Harvey mechanism of anomaly cancellation.\\

\end{enumerate}

The rest of the paper is organized as follows. In Sec. II, we demonstrate the $Z_2$ index associated with the generalized Jackiw-Rossi Hamiltonian. In Sec. III, we explicitly present the mapping of the three-dimensional Dirac Hamiltonian with fully gapped pairings to two copies of the generalized Jackiw-Rossi Hamiltonian under appropriate conditions. Section IV is devoted to establish the invariance of zero energy sub-space under the influence of the momentum along the vortex line. In Sec. V, we exemplify this claim by computing the dispersive modes with energies $E=\pm k_z$, as well as the vortex zero energy modes for particular choice of parameters. There we show that the dispersive modes can indeed be constructed from two-dimensional zero-energy modes by performing a perturbation calculation in terms of $k_z$. In Sec. VI, we summarize our main findings regarding the $Z_2$ index of $H^{JR}_{gen}$ and its consequences for superconducting states of three-dimensional Dirac fermions. In addition we propose a generalization of the Callan-Harvey model of axial superfluid and discuss the consequences $Z_2$ index on anomaly equations. Additional details of the derivation of $Z_2$ index for the generalized Jackiw-Rossi Hamiltonian are presented in Appendix A. We relegate the details of the zero-mode calculations at the special points $\lambda=\pm \chi$ to Appendix B. Appendix C contains some details on the exact evaluation of gapless dispersive modes with underlying topological and $s$-wave pairings in three dimensions.      
\\

\section{$Z_2$ index of generalized Jackiw-Rossi Hamiltonian}

Upon setting set $\lambda=\chi=0$ in Eq. (\ref{genJR}), the resulting Hamiltonian conforms to the one studied originally by Jackiw and Rossi for point vortices \cite{JR-original}. It was shown in Ref. \onlinecite{JR-original} and Ref. \onlinecite{weinberg-index} that for arbitrary vorticity $(n)$ of the point vortex, there exists precisely $n$ number of zero energy states. This problem belongs to class BDI in the Altland-Zirnbauer  classification.\cite{Altland, 10foldclassification} The generalized Jackiw-Rossi Hamiltonian in Eq. (\ref{genJR}), which on the other hand belongs to class D, may as well support states at zero energy depending on the vorticity and the relative strength of the parameters $\lambda$ and $\chi$ and the asymptotic value of the order parameter as $r \to \infty$ ($\Delta_0$). Possible existence of the zero energy mode necessarily requires a spectral symmetry of the Hamiltonian, $H^{JR}_{gen}$. Such symmetry of $H^{JR}_{gen}$ is generated by an anti-unitary operator, namely, $A= U K$, where $U$ is the unitary operator and $K$ is complex conjugation \cite{herbut-lu-genJR, seradjeh}. Without any loss of generality, one can commit to a representation in which $\Gamma_1, \Gamma_2$ are real and $\Gamma_3, \Gamma_4$ are imaginary.\cite{okubo, explanationclifford} In that representation, $U$ is the identity operator, and the antilinear operator $A$ is simply the complex conjugation. Furthermore, if there exists any state at precise zero energy in the spectrum of $H^{JR}_{gen}$, it needs to be an eigenstate of $A$ with eigenvalue $\pm 1$.
\\

\subsection{Spectral symmetry and zero modes}
Let us now focus on the zero energy modes of the Hamiltonian $H^{JR}_{gen}$, in Eq. (\ref{genJR}). Since all the \emph{four}-dimensional representations of mutually anticommuting matrices are unitarily equivalent, for our convenience we choose to work with \cite{herbut-juricic-roy}
\begin{eqnarray}\label{genJRmatrices}
&&\Gamma_1=- \sigma_3 \otimes \sigma_1, \: \Gamma_2 =\sigma_0 \otimes \sigma_2, \: \Gamma_3= \sigma_1 \otimes \sigma_1, \nonumber \\
&&\Gamma_4=\sigma_2 \otimes \sigma_1, \: \Gamma_5= \Gamma_1 \Gamma_2 \Gamma_3 \Gamma_4=\sigma_0 \otimes \sigma_3.
\hspace{-1.5cm}
\end{eqnarray}
Let us define a four component spinor as
\begin{equation}\label{genJRspinor}
\psi^\top (\vec{x})= \left( u_1, v_1,u_2,v_2\right)(\vec{x}),
\end{equation}
and here we wish to solve
\begin{equation}\label{zeroequation}
H^{JR}_{gen} \: \psi(\vec{x})=0.
\end{equation}
In this representation, the antilinear operator, which anticommutes with $H^{JR}_{gen}$ and ensures its the spectral symmetry is
\begin{equation}\label{antilinear}
A= i \Gamma_2 \Gamma_3 \; K = \; \left( \sigma_1 \otimes \sigma_3 \right) \; K.
\end{equation}
This antiunitary operator leaves the zero energy sub-space invariant. Hence, the zero energy state needs to be an eigenstate of the operator $A$, with eigenvalue either $+1$ or $-1$, implying
\begin{equation}\label{spinorcomprel}
\left( \begin{array}{c}
u_1 \\ v_1 \\ u_2 \\ v_2
\end{array} \right) (\vec{x})= \pm
\left( \begin{array}{c}
u^*_2 \\ -v^*_2 \\u^*_1 \\- v^*_2
\end{array} \right)(\vec{x}).
\end{equation}
Upon imposing the constraint on the spinor components with the $+$ sign in the last equation, the four coupled differential equations for the zero energy mode reduce to only two, which read as
\begin{eqnarray}\label{eqgenJr}
(-i) e^{i \phi} \bigg( \partial_r + \frac{i}{r} \partial_\phi \bigg) v^*_2 + \Delta_r e^{-i n \phi} v_2 + (\lambda + \chi)u^*_2 &=&0 \nonumber \\
(i) e^{-i \phi} \bigg( \partial_r - \frac{i}{r} \partial_\phi \bigg) u^*_2 + \Delta_r e^{-i n \phi} u_2 - (\lambda - \chi)v^*_2 &=&0. \nonumber \\
\end{eqnarray}
For the ease of calculation, we here chose to work with an underlying antivortex. Our discussion is equally applicable for point vortex. The above set of equations has been derived for $\mathbf{a}=0$. However, the structure of these equations are not qualitatively affected by $\mathbf{a}$, and we will explicitly account for the gauge field in the subsequent sections. From the above equations, the imaginary factor can be removed by redefining the spinor components as
\begin{equation}\label{pi4phase}
v^*_2 (\vec{x}) =e^{i \frac{\pi}{4}} V^* (\vec{x}), \: u^*_2 (\vec{x}) =e^{-i \frac{\pi}{4}} U^* (\vec{x}).
\end{equation}
Next we find the zero-mode solution of the generalized Jackiw-Rossi Hamiltonian, shown in Eq. (\ref{zeroequation}), with underlying point antivortex of even and odd vorticities. One may have taken the zero energy state, as an eigenstate of $A$ with eigenvalue $-1$. However, this corresponds to a phase rotation by $\exp \left(i \frac{\pi}{4} \Gamma_5\right)$, which is not an observable \cite{roy-smith-kennett}. 
\\

\subsection{Even vorticity: $n= 2 s$}

Let us first consider even vortex, i.e., $n= 2 s$, where $s$ is a positive integer. With even vorticity, single-valued solutions of the zero modes can only be found assuming \cite{ghaemi-wilczek, roy-smith-kennett}
\begin{eqnarray}\label{multiangans} \hspace{-0.2cm}
V (\vec{x}) &=& e^{i l \phi} f_1 (r) + e^{i m \phi} g_1 (r), \nonumber \\
U (\vec{x}) &=& e^{i p \phi} f_2 (r) + e^{i q \phi} g_2 (r),
\end{eqnarray}
where $l,m,p,q$ are restricted to be integers. The consistent solutions of Eq. (\ref{zeroequation}), with the above ansatz can be obtained upon imposing the following constraints over the angular momenta:
\begin{equation}\label{evenvortcont}
l+m=2s+1; p+q = 2 s -1; p=l-1; q=2 s -l,
\end{equation}
which we obtain here by matching the phase factors in Eq.(\ref{eqgenJr}). One should notice that the above constraints among different angular momenta channels of $V$ and $U$ arise only when $\lambda$ and/or $\chi$ are nonzero. Otherwise, $U$ and $V$ are decoupled from each other, same as in the original work by Jackiw and Rossi \cite{JR-original}. 
\\

Finding closed form solutions of the radial functions in Eq.~(\ref{multiangans}) is an involving task. Nevertheless, possible existence of the zero modes can be established by studying the asymptotic structure of these functions. We here only present the final outcome of our analysis. Additional details of this calculation can be found in Appendix A. In the vicinity of the origin ($r \rightarrow 0$), as $\Delta_r \rightarrow 0$ 
\begin{eqnarray}\label{r0f1solEmain}
f^<_1(r)=C_l \; J_{\sqrt{l}} \big[ r \; \zeta  \big], \: \: g^<_1(r)=\tilde{C}_l \; J_{\sqrt{2s-l+1}} \big[ r \; \zeta  \big],
\end{eqnarray}
if $\lambda>\chi$, whereas   
\begin{eqnarray}\label{rof2solEmain}
f^<_1(r)=C_l \; I_{\sqrt{l}} \big[ r \; \zeta  \big], \: \: g^<_1(r)=\tilde{C}_l \; I_{\sqrt{2s-l+1}} \big[ r \; \zeta  \big],
\end{eqnarray}
for $\lambda<\chi$, with a constraint $l=0,1, \cdots, (2 s-1)$. Therefore, with even vorticity ($n=2 s$), there may exist $n$ number of zero energy states. Here, we denote all the radial functions in the vicinity of the origin as $X^<(r)$, where $X=f_1,f_2,g_1,g_2$. $C_l$, $\tilde{C}_l$ are arbitrary constants, and $\zeta=\sqrt{\lambda^2-\chi^2}$. $J_k$ and $I_k$ are, respectively, the Bessel and the modified Bessel functions of first kind of order $k$. The remaining two functions for small $r$ go as
\begin{eqnarray}\label{f2g2relationr0}
f^<_2(r) &=& - \bigg( \frac{1}{\lambda+\chi}\bigg) \: \bigg( \partial_r + \frac{l}{r}\bigg) f^<_1(r), \nonumber \\
g^<_2(r) &=& -\bigg( \frac{1}{\lambda+\chi} \bigg) \: \bigg( \partial_r + \frac{2 s+1-l}{r}\bigg) g^<_1(r). 
\end{eqnarray}
Therefore all the four radial functions close to the origin is defined in terms of \emph{one}(1) arbitrary constant. \\

Far from the origin ($r \rightarrow \infty$), as $\Delta_r \rightarrow \Delta_0$(constant), these functions take the form 
\begin{eqnarray}\label{fginfCI}
\label{f1infCI} f^>_1(r)=g^>_1(r)=\frac{{\cal E}_1}{2} \bigg[ C_{+} \exp \left (\alpha^F_+ r \right) + C_{-} \exp \left (\alpha^F_- r \right)  \bigg], \nonumber \\
\label{f2infCI} f^>_2(r)=g^>_2(r)=\frac{{\cal E}_2}{2} \bigg[ C_{+} \exp \left (\alpha^F_+ r \right) -C_{-} \exp \left (\alpha^F_- r \right)  \bigg], \nonumber \\
\end{eqnarray}
if $\chi^2 < \lambda^2$ and consequently $\chi^2-\lambda^2<\Delta^2_0$. $C_\pm$ are arbitrary constants, ${\cal E}_1= 1$, ${\cal E}_2 =\sqrt{(\chi-\lambda)/(\chi+\lambda)}$, and $\alpha^F_\pm=-\Delta_0 \pm \sqrt{\chi^2-\lambda^2}$. The radial functions far away from the origin are denoted by $X^>(r)$, where $X=f_1,f_2,g_1,g_2$. If, on the other hand, $\chi^2-\lambda^2 >\Delta^2_0$,
\begin{eqnarray}\label{fginfCII}
f^>_j(r)=\frac{{\cal E}_j}{2} \bigg[ C_{+} \exp \left (\alpha^F_+ r \right) -(-1)^{j} \tilde{C}_{-} \exp \left (\alpha^F_- r \right)  \bigg], \nonumber \\
g^>_j(r)=\frac{{\cal E}_j}{2} \bigg[ C_{+} \exp \left (\alpha^F_+ r \right) +(-1)^{j} \tilde{C}_{-} \exp \left (\alpha^F_- r \right)  \bigg],
\end{eqnarray}
for $j=1,2$, where $\tilde{C}_-$ is also a arbitrary constant and $\alpha^G_\pm=\Delta_0 \pm \sqrt{\chi^2-\lambda^2}$. Therefore, irrespective of the mutual strength of $\lambda, \chi$ and $\Delta_0$, the radial part of the zero mode solutions far away from the origin is always defined by \emph{two} (2) arbitrary constants.
\\

\emph{Boundary conditions}: To obtain self-consistent solutions of the zero energy states, the spinor components need to satisfy the boundary conditions at a particular point $r=\xi$(say). Since the solutions in the asymptotic regions are obtained by solving a second-order differential equations, for each functions, we need to match values and the first derivatives of $X^<(r)$ and $X^>(r)$ at $r=\xi$, where $X=f_1,g_1,f_2,g_2$. However, for example, when we impose these boundary conditions over $f_2(r)$, they remove the arbitrariness from two out of the three constants, either $\left(C_+,C_-, C_l \right)$ or $\left(C_-,\tilde{C}_-, C_l \right)$, depending on whether $\chi^2-\lambda^2<0$ or $\chi^2-\lambda^2 >\Delta^2_0$, respectively. In either situation, such elimination immediately removes the arbitrariness of the constants in the solution of $g^>_1(r)$ as well. Therefore, with two fixed constants in $g^>_1(r)$ and one arbitrary constant in the definition of $g^<_1(r)$, it is \emph{impossible} to satisfy two of its boundary conditions. Similarly, upon imposing the boundary conditions over $f_1(r)$, two out of either $\left( C_+,C_-, C_l\right)$ or $\left(C_-,\tilde{C}_-, C_l \right)$ get fixed, depending on $\chi^2-\lambda^2<0$ or $\chi^2-\lambda^2 >\Delta^2_0$, respectively. In terms of those fixed constants it is once again impossible to satisfy both the boundary conditions for $g_2(r)$. Hence one can conclude that \emph{ for generic parameters} $(\lambda \neq \pm \chi)$, \emph{there exists no zero energy mode for $H^{JR}_{gen}$, when the vorticity is even}.
\\

\subsection{Odd vorticity: $n=2 s +1$}

Next we consider point vortex with odd vorticity, and take $n= 2 s +1$, where $s \geq 0$ is an integer. The single valuedness and the consistency of the zero-mode solutions immediately imply that out of $2 s+1$ possible choices for the zero modes, there is precisely one solution for which we can choose
\begin{equation}\label{singleangans}
V (\vec{x})= e^{i l \phi} f(r); \: \: \: U(\vec{x})= e^{i p \phi} g(r),
\end{equation}
and we obtain the following relation between the angular momenta $l$ and $p$ as
\begin{equation}\label{singleangcont}
l=s+1=p+1,
\end{equation}
from Eqs.~(\ref{zeroequation}) and (\ref{eqgenJr}). Even though we can exactly solve these two coupled differential equations, it is worth analyzing the existence of the zero-mode solutions with the above ansatz, from their asymptotic behaviors. Later, we will present the complete solution and show that these two approaches complement each other. In the vicinity of the origin $(r \to 0)$, upon dropping the contribution from $\Delta_r$, we find that 
\begin{equation}
f^<(r)=f^<_1(r), \quad g^<(r)=-\frac{1}{\lambda+\chi}(\partial_r+ \frac{l}{r}) f^<(r).
\end{equation}
On the other hand, far from the origin ($r \rightarrow \infty$), the solutions are 
\begin{eqnarray}\label{fginfsol}
g^>(r) &=& {\cal E}_2 \sum_{\sigma=\pm} \sigma C_\sigma Q_{\sigma}(\Delta_0,\chi, \lambda, r), \nonumber \\
f^>(r) &=& \sum_{\sigma=\pm} C_{\sigma}Q_{\sigma}(\Delta_0,\chi, \lambda, r),
\end{eqnarray}
only when $\chi^2<\lambda^2$, where 
\begin{equation}
Q_{\sigma}(\Delta_0,\chi, \lambda, r)= \exp \left (-\Delta_0 r+i \sigma \sqrt{\lambda^2-\chi^2}\right). 
\end{equation}
Therefore, upon imposing the boundary conditions, we mentioned in the previous subsection (say at $r=\xi$), two out of three arbitrary constants for each function get fixed, while the remaining one is then set by the normalization condition. If the parameters are such that $\Delta^2_0+\lambda^2 < \chi^2$, there is only one exponentially decaying solution, $\propto Q_-$. Therefore, each function is defined in terms of only \emph{one} arbitrary constant at large distances. Therefore, we have two arbitrary constants and two matching conditions to satisfy. After satisfying the matching conditions, there is no more arbitrary constant is left to set the overall normalization. Hence we can have at least one normalizable zero mode of $H^{JR}_{gen}$ for odd vorticity and 
\begin{equation}\label{solcondmain}
\Delta^2_0 + \lambda^2 > \chi^2.
\end{equation}
Next, we show that this condition can also be obtained from the exact solution of the zero energy mode.
\\

The exact solution of the zero energy states can be found upon assuming
\begin{equation}\label{exactsolans}
f(r)=\tilde{f}(r) e^{-\int^r_0 \Delta_{r'} dr'}, \: \: \: g(r)=\tilde{g}(r) e^{-\int^r_0 \Delta_{r'} dr'}.
\end{equation}
The solution of $\tilde{f}(r)$ for arbitrary $r$ is identical to $f^{<}(r)$, and therefore to $f^<_1(r)$, whereas $g(r)=-(\lambda+\chi)^{-1}(\partial_r+ l/r) f(r)$. However, at large distances, the modified Bessel functions grow exponentially:
\begin{equation}\label{Besselinfr}
I_{\sqrt{l}} (a r) \propto \frac{e^{a r}}{r \sqrt{a}}.
\end{equation}
Therefore, normalizable zero energy modes can only be found when $\Delta^2_0 + \lambda^2 > \chi^2$, identical to the one we found in Eq. (\ref{solcondmain}) by analyzing the asymptotic solutions. On the other hand, when $\lambda > \chi$, $\tilde{f}(r)$ is defined in terms of the Bessel functions of the first kind ($J_k$). The condition in Eq. (\ref{solcondmain}) is then trivially satisfied and we always find a normalizable zero mode.
\\

Besides the above zero-mode solution, there are additional $2 s$ possible ansatz similar to that in Eq. (\ref{multiangans}). The angular momenta satisfy the following constraints:
\begin{equation}\label{multangcontodd}
l+m = 2 s +2, p+q = 2 s , p=l-1, q= (2 s +1) -l,
\end{equation}
once again obtained from Eqs. (\ref{zeroequation}) and (\ref{eqgenJr}). Upon imposing this set of constraints over the angular momenta, one set of coupled differential equations for the functions $f^<_1(r)$ and $f^<_2(r)$, in the vicinity of the origin assumes the identical form as before with an even vortex [see Eq. (\ref{r0f1f2})]. Their solution can readily be found from Eqs. (\ref{r0f1solEmain})-(\ref{f2g2relationr0}), and the remaining two radial functions are 
\begin{eqnarray}\label{g1oddr0sol}
g^<_1(r) = \left\{
\begin{array}{rl}
\tilde{C}_l \; J_{\sqrt{2 s +2 -l}} \big[ r\sqrt{\lambda^2-\chi^2} \big] & \text{if} \: \: \lambda > \chi, \\
\tilde{C}_l \; I_{\sqrt{2 s +2 -l}} \big[ r\sqrt{\chi^2-\lambda^2} \big] & \text{if} \: \: \lambda < \chi,
\end{array} \right. \\
g^<_2(r)=-\frac{1}{(\lambda+\chi)}\left( \partial_r + \frac{2 s +2 -l}{r} \right) g^<_1(r).
\end{eqnarray}
Hence, at small distances there is one arbitrary constant for each of $f_1(r),f_2(r),g_1(r)$, and $g_2(r)$. On the other hand, we have shown in the previous subsection that the large $r$ behavior is independent of the angular momenta. Hence, far away from the origin the radial dependence is captured by Eq. (\ref{fginfCI}) if $ \chi^2-\lambda^2 <0$ and $ \chi^2-\lambda^2 <\Delta^2_0$ or Eq. (\ref{fginfCII}) if $\chi^2-\lambda^2 > \Delta^2_0$. Therefore following the discussion on the matching conditions at $r=\xi$, in the previous subsection, we can argue that there is no zero energy mode with multiple angular momenta ansatz, even when the vorticity is odd.
\\

\subsection{Effects of gauge potential}

In the above derivation, we have neglected the orbital effects of the gauge potential. It can be introduced, for example, considering a simple profile of the magnetic field. Let us assume, that the magnetic field (applied in the $z$ direction, perpendicular to the plane of the vortex) is finite and constant only within a distance $r \leq \xi$, and vanishes for $r > \xi$ \cite{herbut-lu-genJR, melikyan}. Then in the symmetric gauge, one can choose $a_r=0$ and
\begin{equation}\label{aprofile}
a_\phi=\left\{ 
\begin{array}{r l}
 \frac{r}{2 \xi^2} \: \: \mbox{when} \: \: r \leq \xi, \\ \\
\frac{1}{2r} \: \:  \mbox{for} \: \: r>\xi.
\end{array} \right.
\end{equation}
With such a profile of the gauge potential, the ultraviolet and the infrared asymptotic behaviors of all the functions ($f_1,f_2,g_1,g_2, f,g$) remain unchanged. The only significant effect will be at intermediate distance $r \sim \xi$. Hence, our derivation for the $Z_2$ index for the generalized Jackiw-Rossi Hamiltonian remains valid, even in the presence of the gauge field. To further demonstrate this assertion, we explicitly solve the vortex zero mode with this profile of the gauge field in Sec. V.

\subsection{Statement of $Z_2$ index}

After going through the above arguments, we can formally present the statement of the $Z_2$ index associated with the zero modes for the generalized Jackiw-Rossi Hamiltonian. For generic values of $\lambda \neq \pm \chi$, which satisfy $\Delta_{0}^2+\lambda^2>\chi^2$, there exists one normalizable zero mode when the vorticity is \emph{odd}, while with even vorticity all the states are placed at finite energies, in accordance with Eq. (\ref{indexstate}). On the other hand, there are no normalizable zero modes for any vorticity, if $\Delta_{0}^2+\lambda^2<\chi^2$. We notice that the normalizability at infinity is governed by the uniform $H_{JR}^{gen}$. There is a topological phase transition or band inversion at $\Delta_{0}^2+\lambda^2=\chi^2$, and this is the reason for the absence of the normalizable zero energy state on one side.\\

However, there is an exception to the existence of the $Z_2$ index at $\lambda =\pm \chi$ \cite{santos-chamon}. For these special values, the condition  $\Delta_{0}^2+\lambda^2>\chi^2$ is trivially satisfied, which guarantees the normalizability of the zero modes. In the above calculation, we have assumed an antivortex configuration of the order parameter ($\vec{\Delta}$). With an underlying antivortex it can be shown that there exist precisely $n$ number of zero energy states for arbitrary $n$, when $\lambda=-\chi$. Moreover, all the zero energy states are eigenstates of the chirality operator $\Gamma_5$, defined in Eq. (8), with eigenvalue $+1$. On the other hand, for $\lambda=+\chi$, once again we recover the $Z_2$ index for the generalized Jackiw-Rossi Hamiltonian. One can achieve the Hamiltonian describing a point vortex defect by unitarility rotating $H_{JR}$ in Eq. (\ref{JRintro}) by $i \Gamma_4 \Gamma_5$, which changes the relative sign between $\Delta_1$ and $\Delta_2$. When we perform the same operation on the generalized Jackiw-Rossi Hamiltonian in Eq. (\ref{genJR}), it changes the sign of the U(1) gauge field $\mathbf{a} \rightarrow -\mathbf{a}$, and takes $\lambda \rightarrow -\lambda$. Hence, $n$ number of zero energy modes with underlying point vortex appears in the spectrum when $\lambda=+\chi$, whereas the $Z_2$ index remains unaltered for $\lambda=-\chi$. With an underlying vortex when $\lambda=+\chi$, all the $n$ number of zero modes are eigenstate of $\Gamma_5$, however, with eigenvalue $-1$. For the detail solutions at this special values of two parameters $\lambda$ and $\chi$, readers are referred to Appendix B.
 \\

\section{Three-dimensional massive Dirac Hamiltonian with gapped pairings}

Next we focus on momentum independent, time reversal symmetric, gapped paired states of the three-dimensional Dirac fermions. In three spatial dimensions, the Dirac quasiparticles can pair into two fully gapped superconducting states. One of them is trivial $s$-wave pairing, whereas the other one is parity-odd and topologically nontrivial. To study the nature of these paired states in the mixed/vortex phase, let us define an eight component Nambu-Dirac spinor as $\Psi = \left[ \Psi^\top_p (+ \vec{k}),\Psi^\top_h (-\vec{k}) \right]$, where $\Psi^\top_p(+\vec{k})=\Psi^\top(\vec{k})$ and $\Psi^\top_h(-\vec{k})=\Psi_p(\vec{k})$, otherwise
\begin{equation} \label{4compspinor}
\Psi^\top(\vec{k})= \left[ c^+_\uparrow (\vec{k}), c^+_\downarrow (\vec{k}), c^-_\uparrow (\vec{k}), c^-_\downarrow (\vec{k}) \right].
\end{equation}
$c^\pm_s$ corresponds to the annihilation operators for the even and the odd parity states, respectively, with the spin projections $s=\uparrow, \downarrow$. The three-dimensional massive Dirac Hamiltonian in the presence of trivial s-wave ($\Delta_s$) and odd-parity topological ($\Delta_T$) parings takes the form
\begin{equation}\label{DiracHamil}
H_D= \sum_{\vec{k}} \Psi^\dagger (\vec{k})H_{gen} \left[\Delta_s, \Delta_T \right] \Psi(\vec{k}).
\end{equation}
In order to preserve the time-reversal symmetry we do not choose any relative phase between $\Delta_s$ and $\Delta_T$ \cite{goswami-roy-axion}. In the announced eight-dimensional Nambu-Dirac basis ($\Psi$), $H_{gen}\left[\Delta_s, \Delta_T \right]$ takes the form
\begin{widetext}
\begin{eqnarray}\label{totalgenHamil}
H_{gen}\left[ \Delta_s,\Delta_T \right] &=& ( \tau_0 \otimes \alpha_1 ) (k_x -\tau_3 \otimes I_4 a_x) + (\tau_3 \otimes \alpha_2) ( k_y-\tau_3 \otimes I_4 a_y) + (\tau_0 \otimes \alpha_3) ( k_z-\tau_3 \otimes I_4 a_z)  + (\tau_3 \otimes I_4) \mu  \nonumber \\
&+& (\tau_3 \otimes \beta) m_k- (\tau_3 \otimes i \alpha_1 \alpha_2) h_+ - (\tau_3 \otimes i \alpha_3 \Gamma) h_-
- (\tau_2 \otimes \alpha_2 )\Delta^R_T + (\tau_1 \otimes \alpha_2 )\Delta^I_T \nonumber \\
&-& (\tau_2 \otimes i \alpha_1 \alpha_3 )\Delta^R_s + (\tau_1 \otimes i \alpha_1 \alpha_3)\Delta^I_s,
\end{eqnarray}
\end{widetext}
where $\mathbf{a}$ is the electromagnetic gauge potential, and the complex pairing order parameters are defined as
\begin{equation}\label{vortOPn}
\vec{\Delta}_x = \left( \Delta^R_x, \Delta^I_x\right)= |\Delta_x| \left(\cos{n \phi}, \sin{n \phi} \right),
\end{equation}
with $x=T,s$, and $n$ is an integer. Then $n$ counts the vorticity and the Hamiltonians $H_{gen}\left[ \Delta_s,0\right]$ and $H_{gen}\left[ 0,\Delta_T\right]$, respectively, correspond to \emph{line-vortex} defect with underlying s-wave and topologically nontrivial odd parity pairing. Here, $h_\pm$ are respectively the symmetric and the antisymmetric combinations of the Zeeman couplings of the even ($h_1$) and the odd ($h_2$) parity bands, namely,
\begin{equation}\label{zeemancouplings}
h_\pm= \frac{1}{2} \big| h_1 \pm h_2   \big|,
\end{equation}
$\mu$ is the chemical potential, and $m_k$ is the Dirac mass. In what follows, we set $m_k=m=$ constant. The four-dimensional Hermitian matrices are defined as
\begin{equation}
\alpha_1=\left(\begin{array}{c c}
0 & \sigma_1 \\
\sigma_1 & 0\\
\end{array} \right),
\alpha_2=\left(\begin{array}{c c}
0 & \sigma_2 \\
\sigma_2 & 0\\
\end{array} \right),
\alpha_3=\left(\begin{array}{c c}
0 & \sigma_3 \\
\sigma_3 & 0\\
\end{array} \right), \nonumber
\end{equation}
\begin{equation}\label{relativisticmatrices}
\beta=\left(\begin{array}{c c}
\sigma_0 & 0 \\
0 & -\sigma_0\\
\end{array} \right),
\Gamma=\left(\begin{array}{c c}
0 & -i \sigma_0 \\
i \sigma_0 & 0\\
\end{array} \right),
\end{equation}
which together complete the Clifford algebra of five mutually anticommuting four-dimensional matrices. Here, $\sigma_0$ is the two-dimensional unit matrix and $ \vec{\sigma}$ are the Pauli matrices. The other set of two-dimensional matrices, $\left\{ \tau_0, \vec{\tau} \right\}$, operate on the Nambu's index. Here, we have ignored the anisotropy in the Fermi velocity arising from the underlying crystallographic structure, and set $v_x=v_y=v_z=v=1$ \cite{modelhamiltonian-TI}. Next, we cast the pairing Hamiltonians in the $k_z=0$ plane with underlying point vortex defects of the topologically nontrivial odd parity pairing as orthogonal sum of two copies of the generalized Jackiw-Rossi Hamiltonian, under generic situation. Such mapping is shown to be true for the s-wave pairing, however, only if there is no chiral symmetry breaking perturbations, e.g., $m, h_-$.

\subsection{Odd-parity topological pairing}

To perform the above mentioned exercise, it is worth redefining the eight-component Nambu-Dirac spinor as
\begin{eqnarray}\label{topologicalspinor}
\Psi^\top_t=\big[ c^{+}_{\uparrow},c^{-}_{\downarrow},(c^{+}_{\uparrow})^{\dagger},(c^{-}_{\downarrow})^{\dagger},c^{+}_{\downarrow},c^{-}_{\uparrow}, (c^{+}_{\downarrow})^{\dagger},(c^{-}_{\uparrow})^{\dagger} \big] (\vec{x}).
\end{eqnarray}
In this new basis, the part of the Hamiltonian, $H_{gen} \left[ 0,\Delta_T \right]$ describing the point vortex in the $xy$ plane (i.e., with $k_z=0$), is completely block-diagonal, whereas the $k_z$ part is block off-diagonal. For simplicity, let us set all the orbital components of the gauge potential to zero. The total Hamiltonian with only the topological paring then takes the form
\begin{equation}\label{reducedtopoHamil}
H_{gen}\left[ 0, \Delta_T \right] \to H^{vor}_T= \bigg( H^{uL}_T \oplus H^{dR}_T \bigg)+ k_z {\cal M}^T_z,
\end{equation}
where
\begin{eqnarray}\label{topoupper}
H^{uL}_T &=&\gamma_5 \alpha_1  k_x + \beta \gamma_5 \alpha_2 k_y + i \beta \alpha_2 \Delta^R_T+ \alpha_2 \Delta^I_T \nonumber \\
&+& \beta \alpha_3 \gamma_5 \left( m+ h_+\right) + \beta \left( \mu + h_- \right),
\end{eqnarray}
and
\begin{eqnarray}\label{topolower}
H^{dR}_T &=& \gamma_5 \alpha_1  k_x - \beta \gamma_5 \alpha_2 k_y - i \beta \alpha_2 \Delta^R_T+ \alpha_2 \Delta^I_T \nonumber \\
&+& \beta \alpha_3 \gamma_5 \left( m - h_+\right) + \beta \left( \mu - h_- \right).
\end{eqnarray}
The matrix multiplying $k_z$ is ${\cal M}^T_z=\sigma_2 \otimes ( i \alpha_1 \alpha_3)$. Here, we have defined as new matrix
\begin{equation}\label{gamma5}
\gamma_5=\left(\begin{array}{c c}
0 & \sigma_0 \\
\sigma_0 & 0\\
\end{array} \right),
\end{equation}
which anticommutes with $\beta$ and $\Gamma$, but commutes with $\alpha_1, \alpha_2,\alpha_3$. Both $H^{uL/dR}_T$ in Eq. (\ref{topoupper}), and Eq. (\ref{topolower}) assume the form of the generalized Jackiw-Rossi Hamiltonian, shown in Eq. (\ref{genJR}). We need the following identification of the parameters $\chi \equiv m+h_+, \lambda \equiv \mu+h_-$ for $H^{uL}_T$, and $\chi \equiv m-h_+, \lambda \equiv \mu-h_-$ for $H^{dR}_T$.
\\

Therefore, with a point vortex defect of underlying odd-parity topological paring one ends up with \emph{two} Majorana modes, in the presence of generic perturbations. However, these Majorana modes can only be found when the vorticity of the point vortex is \emph{odd}, as we have shown in the previous section. Since the magnetic field gets screened beyond the core of the vortex, the Zeeman couplings $h_\pm$ are finite only within the vortex core. Hence two normalizable Majorana modes can be achieved only when
\begin{equation}
\Delta^2_0 + \mu^2> m^2.
\end{equation}
\\

\subsection{$s$-wave pairing}

A similar exercise can also be performed with an underlying s-wave pairing, which also assumes the form of the generalized Jackiw-Rossi Hamiltonian, when $m=0$ and $h_-=0$. Let us now rewrite $H_{gen}\left[ \Delta_s,0 \right]$ in the basis
\begin{eqnarray}\label{spinorswave}
\Psi^\top_s= \big[ c^+_\uparrow, c^-_\downarrow, (c^+_\downarrow)^\dagger,(c^-_\uparrow)^\dagger,
c^+_\downarrow, c^-_\uparrow, (c^+_\uparrow)^\dagger,(c^-_\downarrow)^\dagger \big](\vec{x}).
\end{eqnarray}
The eight-dimensional Hamiltonian, $H_{gen}\left[\Delta_s,0 \right]$ then takes the form
\begin{equation}\label{reducedswave}
H_{gen}\left[ \Delta_s,0 \right] \to H^{vor}_s= \bigg( H^{uL}_s \oplus H^{dR}_s \bigg) + k_z {\cal M}^s_z,
\end{equation}
similar to the Eq. (\ref{reducedtopoHamil}). The diagonal blocks of $H^{vor}_s$ are
\begin{eqnarray}\label{swaveupper}
H^{uL}_s &=& \gamma_5 \left( \alpha_1 k_x + \alpha_2 k_y \right) + \Delta^R_s \alpha_3 + \Delta^I_s i \alpha_3 \beta + \mu \beta \nonumber \\
&+& \gamma_5 \alpha_3 h_+ - i \beta \alpha_1 \alpha_2 m + I_4 h_-,
\end{eqnarray}
and
\begin{eqnarray}\label{sawavelower}
H^{dR}_s &=& \gamma_5 \left( \alpha_1 k_x - \alpha_2 k_y \right) - \Delta^R_s \alpha_3 - \Delta^I_s i \alpha_3 \beta + \mu \beta \nonumber \\
&-& \gamma_5 \alpha_3 h_+ - i \beta \alpha_1 \alpha_2 m - I_4 h_-.
\end{eqnarray}
Here as well the $k_z$ appears as block off-diagonal element, and the matrix multiplying $k_z$ is ${\cal M}^s_z =\sigma_2 \otimes (i \alpha_1 \alpha_3)$, identical to ${\cal M}^T_z$. Therefore, in the absence of the Dirac mass ($m$) and the antisymmetric combination of the Zeeman coupling ($h_-$), both $H^{uL}_s$ and $H^{dR}_s$ are equivalent to the generalized Jackiw-Rossi Hamiltonian, shown in Eq. (\ref{genJR}), where $\lambda= \mu$, $\chi=h_+$ for both $H^{uL/dR}_{gen}$. Hence, in the absence of any chiral symmetry breaking perturbations, the point vortex of underlying s-wave order can also support two Majorana zero modes, when the vorticity is odd. The zero energy modes in the absence of $\mu, h_+$, are also the eigenstates of $\beta \gamma_5=-\sigma_2 \otimes \sigma_0$, with definite chirality. This matrix defines the chirality of massless three-dimensional Dirac fermions, and should not be confused with $\Gamma_5=\sigma_0 \otimes \sigma_3=i\alpha_2\alpha_1$, which defines the chirality of the original Jackiw-Rossi Hamiltonian in Eq.~(\ref{JRintro}). Any chiral symmetry breaking perturbation of the three-dimensional Dirac fermions, such as Dirac mass ($m$) and $h_-$ cause mixing among these two states and the spectrum becomes \emph{gapped}.
\\

\section{Perturbation theory for line vortex about $k_z$}

In this section we show that the $Z_2$ index for the fermionic zero modes bound to the point vortex, also dictates the number of one-dimensional dispersive modes along the line vortex in three spatial dimensions. The momentum along the vortex core ($k_z$) is shown to leave the zero energy sub-space of the underlying two-dimensional Hamiltonian (multiple copies of $H_{gen}^{JR}$) invariant. Consequently, a first-order perturbation theory for $k_z$ is exact and the dispersive modes along the vortex core are the linear combinations of the vortex zero modes. We here prove this statement with underlying topologically non-trivial odd-parity and s-wave pairings separately.
\\

\subsection{Topological pairing}

Let us first consider the generalized Dirac Hamiltonian with the topological pairing, $H_{gen} \left[0, \Delta_T \right]$. We have shown in the previous section that $H_{gen} \left[ 0,\Delta_T \right]$ is unitarily equivalent to two copies of the generalized Jackiw-Rossi Hamiltonian, when $k_z=0$. Therefore, $H_{gen} \left[ 0, \Delta_T ; k_z=0\right]$ hosts two Majorana zero modes for odd vorticity, under generic situation. These two zero energy states constitute a two-dimensional basis, which remains invariant by any operator that commutes or anticommutes with the Hamiltonian. If we turn off all the perturbations, namely $m$, $\mu$, $h_+$ and $h_-$, then there are \emph{four} such candidates falling into the second category. Together they close a $Cl(3) \times U(1)$ algebra. The three mutually anticommuting matrices, closing the $Cl(3)$ subalgebra act like standard two-dimensional Pauli matrices. The remaining one, belonging to the U(1) commutes with three matrices, which close the $Cl(3)$ sub-algebra. With underlying topological pairing the $Cl(3) \times U(1)$ algebra is constituted by
\begin{equation}\label{cl3topo}
\vec{M}_T = \left\{ \tau_0 \otimes \alpha_3, \tau_0 \otimes \beta, \tau_0 \otimes \Gamma, \tau_0 \otimes i \alpha_1 \alpha_2 \right\},
\end{equation}
where the last entry belongs to the U(1) part. However, due to Nambu's particle-hole doubling of the original problem, an 8-dimensional $k$-dependent perturbation, $ a_k M_k$ can acquire a finite expectation value, only if the matrix $M_k$ satisfies the algebraic constraint
\begin{equation}\label{massmatriconst}
M_k = \mp \left( \tau_1 \otimes I_4 \right) \; M^\top_k \; \left( \tau_1 \otimes I_4 \right).
\end{equation}
In the above equation, the $\mp$ signs depend on whether the coefficient $a_k$ is even or odd under the parity transformation ($\mathbf{k} \rightarrow - \mathbf{k}$). If the coefficient $a_k$ is of even parity, the only matrix satisfying the above constraint is $\tau_0 \otimes \Gamma$. This matrix represents a {\em parity and time-reversal odd Dirac mass} ($m_{PT}$). On the other hand, for an odd parity $a_k$ (e.g., linear in $\mathbf{k}$) the remaining three matrices can acquire finite expectation values from the zero energy sub-space. Notice that one of the matrices, $\tau_0 \otimes \alpha_3$, appearing in the $Cl(3)$ part of $\vec{M}_T$, multiplies $k_z$ in $H_{gen}$, shown in Eq. (\ref{totalgenHamil}). Therefore, the momentum along the vortex core $k_z$ does not cause any mixing of the zero energy states with the rest of the spectrum. Consequently, a first order perturbation calculation in terms of $k_z$ leads to the \emph{exact} solution of the dispersive modes along the vortex core. The matrix $\tau_0 \otimes \alpha_3$ acts as an off-diagonal Pauli matrix in the zero energy subspace, and hence the one-dimensional dispersive modes are the symmetric and the antisymmetric combinations of the fermionic zero modes with underlying point vortex. The energies of these two dispersive modes are $E= \pm k_z$. When $k_z=0$, we have two Majorana fermions, which hybridize via $k_z$ and become complex fermions.
\\

The exactness of the perturbation theory in terms of $k_z$ is also applicable when we take into account the perturbations $m,\mu,h_\pm$. However, not all the matrices in $\vec{M}_T$ anticommute with the generic Hamiltonian $H_{gen}\left[ 0,\Delta_T ; k_z=0\right]$. Before, we proceed to prove this statement it is worth appreciating an algebraic identity.\cite{wijewardhana, herbut-neel, roy-realaxial-zeromode} Expectation value of an operator (${\cal M}$) can be expressed as
\begin{equation}\label{expectationvalue}
\langle {\cal M } \rangle = \frac{1}{2} \bigg( \sum_{occupied}- \sum_{empty} \bigg) \Psi^\dagger_E {\cal M} \Psi_E,
\end{equation}
where $\Psi_E$ are the eigenstates of a generic Hamiltonian, ${\cal H}$. If there exists a matrix, say $T$, which anticommutes with ${\cal H}$ and commutes with ${\cal M}$, the above mentioned sum is restricted to the zero energy subspace. When $m=\mu=h_\pm=0$, and ${\cal M} = (\tau_0 \otimes \alpha_3 ) k_z$, one can choose $T=\tau_0 \otimes i \alpha_1 \alpha_2$. When $m,\mu,h_\pm$ are finite, we can still find a matrix, $T$ with requisite criteria, for the chosen ${\cal M} = ( \tau_0 \times \alpha_3 ) k_z$. If all the perturbations $m,\mu,h_\pm$ are nonzero, we cannot find any unitary matrix for $T$. Rather there is an antiunitary operator ($A_T$), namely $( \tau_1 \otimes i \beta \alpha_1 \alpha_2)\; $K, where K is the complex conjugation, which anticommutes with the Hamiltonian $H_{gen}\left[0,\Delta_T;k_z=0 \right]$ and commutes with $( \tau_0 \otimes \alpha_3 ) k_z$. Therefore, we can choose  $T= A_T$. Hence, the one-dimensional dispersive modes are always the symmetric and the antisymmetric combinations of the zero energy modes bound to the point vortex. Even in the presence of the gauge fields, we can still choose $T=A_T$, and the above conclusions remain unaltered.
\\

\subsection{s-wave pairing}

A similar conclusion can be arrived at even with an underlying s-wave pairing if we turn off all the chiral symmetry breaking perturbations, for example, $m,h_-$. Furthermore, when $\mu, h_+$ is set to zero the $Cl(3) \otimes U(1)$ algebra of the matrices, anticommuting with the Hamiltonian $H_{gen}\left[\Delta_s,0; k_z=0 \right]$ is constituted by
\begin{equation}\label{cl3swave}
\vec{M}_s = \left\{ \tau_0 \otimes \Gamma, \tau_0 \otimes \alpha_3 , \tau_3 \otimes \beta, \tau_0 \otimes i \alpha_1 \alpha_2\right\},
\end{equation}
where the last entry belongs to the U(1) part. Appearance of the matrix $\tau_0 \otimes \alpha_3$ in the $Cl(3)$ part of $\vec{M}_s$, immediately guarantees that the dispersive one-dimensional modes can be obtained by performing a perturbative calculation over $k_z$ within the two-dimensional basis spanned by the localized fermionic zero energy modes due to a point vortex in the $k_z=0$ plane. It can also be confirmed from the Eq. (\ref{expectationvalue}), upon choosing $T= \tau_0 \otimes i \alpha_1 \alpha_2$ and ${\cal M}=\tau_0 \otimes \alpha_3$. Otherwise, the matrix $\tau_0\otimes \alpha_3$ acts as the \emph{diagonal} Pauli matrix. Hence the dispersive modes with energies $E=\pm k_z$, are identical to the fermionic zero mode due to the point vortex. Let us now incorporate a finite chemical potential ($\mu$) and the {\em symmetric} Zeeman coupling $h_+$. One can then choose $T= \left( \tau_2 \otimes i \beta \alpha_1 \alpha_2 \right)$ K, for ${\cal M}= ( \tau_0 \otimes \alpha_3 ) k_z$. In conjunction with such choice of $T$, Eq. (\ref{expectationvalue}) guarantees that the dispersive modes with $E=\pm k_z$, are exactly the two fermionic zero modes bound to the point vortex.
\\

It is worth mentioning that one of matrices in the $Cl(3)$ subgroup, namely $\tau_3 \otimes \beta$ appears in $H_{gen}$ in Eq. (\ref{totalgenHamil}) with the Dirac mass $(m)$. This matrix also satisfies the condition in Eq. (\ref{massmatriconst}), if its coefficient is momentum independent or even under $\mathbf{k} \rightarrow - \mathbf{k}$. Therefore, the internal structure of the zero energy sub-space shows that the Dirac mass is sufficient to cause splitting of these two states and place them at finite energies, $\pm m$.
\\

\section{Exact and perturbative solutions of line vortex}

The internal structure of the zero energy modes of the point vortex allowed us to show that the one-dimensional dispersive modes along the core of the vortex can be constructed from the localized Majorana zero modes due to the point vortex. Thus the number of zero energy modes in the spectrum of the two-dimensional generalized Jackiw-Rossi Hamiltonians, $H^{uL}_{T}$ and $H^{dR}_{T}$ with underlying odd-parity topological pairing, or $H^{uL}_{s}$ and $H^{dR}_{s}$ with underlying s-wave pairing, is exactly the number of gapless states along the vortex line. To exemplify this claim, we here first present the exact solutions of the dispersive modes for line-vortex as well as the zero mode solutions of the point vortex, for particular choices of the parameters in $H_{gen} \left[0, \Delta_T \right]$ and $H_{gen} \left[\Delta_s, 0 \right]$. Then we show that the one-dimensional dispersive modes with energies $E=\pm k_z$, are either linear combinations of (for the topological pairing) or exactly (for the s-wave pairing) the zero modes for the point vortex. Some particular limits of this problem has been considered previously in Refs. \onlinecite{nishida, Tagliacozzo, Yuki, Goswami-Roy-effectivetheory}. We here present only the final results. Some additional details of this calculation can be found in Appendix C.
\\

\subsection{Topological pairing with $\mu=h_+=0$}

We consider a line vortex along the $z$ direction with an underlying odd-parity topological pairing. The solution for dispersive mode with energies $E=\pm k_z$, when $\mu=h_+=0$, in the basis of eight-component Nambu-Dirac spinor shown in Eq. (\ref{4compspinor}) reads as
\begin{eqnarray}\label{Epktopo}
|E=\pm k_z \rangle = {\cal R}  (r,z) \left(
\begin{array}{c}
g(r) e^{-i \phi}\\
\pm i f(r)\\
\mp g(r) e^{-i \phi}\\
i f(r)\\
 i g(r) e^{i \phi}\\
\pm f(r)\\
\mp i g(r) e^{i \phi}\\
f(r)
\end{array}
\right),
\end{eqnarray}
where the function ${\cal R}(r,z)$ takes the form 
\begin{equation}
{\cal R}(r,z)=\exp{\left(-i\frac{\pi}{4}+i k_z z-\int^r_0 \Delta_{r'} dr'\right)}.
\end{equation}
Taking the profile of the vector potential $\mathbf{a}$ to be same as in Eq.~(\ref{aprofile}), one obtains
\begin{eqnarray}\label{gfrlxi}
g(r) &=& c_1 \; I_1\bigg[ r \sqrt{m^2-h^2}\bigg], \nonumber \\
f(r) &=& c_1 \sqrt{\frac{m+h}{m-h}}\; I_0\bigg[ r \sqrt{m^2-h^2}\bigg],
\end{eqnarray}
when $r \ll \xi$. Outside the core of the vortex ($r>\xi$),
\begin{eqnarray}\label{gfrgxi}
g(r) &=& c_3 I_{1/2} \bigg[ r |m|\bigg] + c_4 I_{-1/2} \bigg[ r |m|\bigg], \nonumber \\
f(r) &=& c_4 I_{1/2} \bigg[ r |m|\bigg] + c_3 I_{-1/2} \bigg[ r |m|\bigg].
\end{eqnarray}
The solutions and their first derivatives need to be matched at $r=\xi$, where the solutions for $r<\xi$ can be found by replacing $h^2 \rightarrow h^2+1/2\lambda^2$ \cite{herbut-lu-genJR}. It eliminates two out of three arbitrary constants from $f(r)$ and $g(r)$, while the remaining one is fixed by the normalization condition.
\\

On the other hand, solutions for two fermionic zero modes with an underlying point vortex, when $\mu=h_+=0$, are the following:
\begin{equation}\label{vorttoposol1}
| \Psi^0_1 \rangle = {\cal R}(r,0) \left(
\begin{array}{c}
g(r) e^{-i \phi}\\ 0 \\ 0 \\ i f(r) \\ i g(r) e^{i \phi}\\ 0 \\ 0 \\ f(r)
\end{array} \right),
| \Psi^0_2 \rangle ={\cal R}(r,0) \left( \begin{array}{c}
0\\ -i f(r)\\ g(r) e^{-i \phi}\\ 0\\ 0\\ -f(r)\\ i g(r) e^{i \phi}\\ 0 \end{array} \right).
\end{equation}
It is now evident that the solution for two one-dimensional dispersive modes, $| E=+k_z \rangle$ and $| E=-k_z \rangle $ in Eq. (\ref{Epktopo}), are respectively the antisymmetric and the symmetric combination of two fermion zero modes $| \Psi^0_1 \rangle$ and $| \Psi^0_2 \rangle$, shown in Eqs.~(\ref{vorttoposol1}), in the presence of point vortex.
\\

Now we point out some subtleties regarding the perturbative treatment of the $k_z$ term. In the absence of the chemical potential and $h_+$, we have solved the dispersive modes exactly, and the velocity along the $z$ direction remains the same as the original Dirac quasiparticle's Fermi velocity. In this particular case, if we treat the $k_z$ part perturbatively with respect to the zero modes of the planar Hamiltonian, we reproduce the exact results for the eigenspinor and the eigenenergies from the first order degenerate perturbation theory. However, the perturbative treatment of $k_z$ in the presence of $\mu$ does not yield the exact eigenspinor and the eigenspectrum. This mismatch can be attributed to the fact that $[\tau_0\otimes \alpha_3, \tau_3\otimes I_4]=0$. Therefore one needs to solve the eigenproblem of the dispersive modes exactly using numerical methods. This subtlety due to the commuting matrices can also be found in an opposite limit. Consider, solving the dispersive mode exactly in the absence of $\mu$ and $h_+$. The matrix element of $\mu$ (which is a chiral chemical potential for the BdG quasiparticles) in the basis of the obtained dispersive states vanishes. Thus,the simultaneous presence of commuting operators require an exact treatment to avoid fallacious perturbative conclusions.

\subsection{s-wave paring with $m=\mu=h_\pm=0$}

Let us now consider a line vortex along the $z$ direction with underlying s-wave paring. The dispersive mode can be solved analytically when we set $m=\mu=h_\pm=0$. Solution for two dispersive modes along the vortex line then takes the form (see Appendix C 2 for detail)
\begin{eqnarray}\label{swave1D}
|+k_z\rangle = C^+_- { \cal R}(r,z) \left(
\begin{array}{c}
0\\
- 1\\
0\\
1\\
0\\
- i\\
0\\
i
\end{array}
\right),
|-k_z\rangle = C^-_- {\cal R}(r,z) \left(
\begin{array}{c}
0\\
i\\
0\\
i\\
0\\
1\\
0\\
1
\end{array}
\right). \nonumber \\
\end{eqnarray}
Next we proceed to obtain the zero energy ($E=0$) modes with an underlying point vortex of s-wave paring in the $k_z=0$ plane. It can readily be solved from Eq. (\ref{swaveeqn}) upon setting $E=0, k_z=0$. With an underlying s-wave pairing $k_z$ acts like $\sigma_3=\mbox{diag}. (1,-1)$ matrix in the zero energy sub-space. Therefore it does not cause any mixing between two Majorana modes, bound to the point vortex. Consequently, the gapless modes are same as the vortex zero modes, multiplied by the plane-wave factor $\exp{(i k_z z)}$. \\

\section{Summary and Discussions}

In this paper, we have demonstrated that the original Jackiw-Rossi Hamiltonian describing a point vortex defect in two spatial dimensions [see Eq.~(\ref{JRintro})], can be augmented by two additional terms [see Eq.~(\ref{genJR})], which still possesses a spectral symmetry. Consequently, the generalized Jackiw-Rossi Hamiltonian of Eq.~(\ref{genJR}) may support zero modes. For generic values of the perturbation parameters satisfying the condition in Eq.~(\ref{solcondmain}), we obtain a single zero mode only for the odd vorticity. In contrast, there are no zero modes for the even vorticity, for generic perturbations. To demonstrate the emergence of the $Z_2$ index for the zero modes, we have employed a method of matching asymptotic solutions, which correctly captures all the known results within a single framework. We have also found that there exist special values of $\lambda=\pm \chi$, for which there are $n$ number of zero modes respectively for a vortex and an antivortex of vorticity $n$.\\

One of the main goals of this paper is to determine the number of gapless one-dimensional modes along the line vortex of a gapped paired states of three-dimensional Dirac quasiparticles. In order to answer this question, we have mapped the problem into the determination of the number of vortex zero modes of appropriate $H_{gen}^{JR}$'s. Through this procedure, we have succeeded in showing that the number of gapless modes is also dictated by the $Z_2$ index of $H_{gen}^{JR}$'s. We have exemplified this by considering the topological odd parity and trivial s-wave pairings. If the underlying Dirac fermions are massless, then in the absence of the chemical potential and the Zeeman couplings, both types of pairing lead to two copies of appropriate $H_{JR}$ for $k_z=0$. Consequently, the number of the gapless modes is governed by the $Z$ index theorem of Weinberg\cite{JR-original, weinberg-index}. When generic perturbations are considered, only the topological pairing sustains \emph{two} gapless modes for odd vorticity, in accordance with the $Z_2$ index. With an underlying s-wave pairing the vortex zero energy modes and thus the gapless dispersive modes can only be found only when chiral-symmetry breaking perturbations in the particle-hole channel, such as the Dirac mass, $h_-$, are absent. \\

Our calculations for isolated vortex can be extended in perturbation theory set up for multiple vortices in the dilute vortex limit. In the presence of multiple vortices, if we again consider $k_z=0$, there will be tunneling within each copy of $H_{gen}^{JR}$'s [see Eqs. (\ref{topoupper}) and (\ref{topolower})] for topological superconductor. As far as the topology of the order parameter field is concerned, there is no difference between a $n$ vortex and widely separated $n$ number of single vortices. Our analysis now suggests an interesting \emph{even-odd} effect based on the $Z_2$ index. When the net vorticity is odd, the gapless state will survive the tunneling effects. In contrast, the gapless state will be absent for a net even vorticity. This consideration can also be extended to nondilute limit following the strategy in Ref.~\onlinecite{Silaev}. For similar even-odd effect in $p+ip$ superconductor, see Ref. \onlinecite{gurarie-radzihovsky}.\\

With an underlying s-wave pairing, the one-dimensional dispersive modes are gapped in the presence of a Dirac mass ($m$). Hence the one-dimensional Hamiltonian along the vortex core conforms to a massive Dirac Hamiltonian in one spatial dimension. Thus with a domain wall configuration of $m$, the one-dimensional Hamiltonian corresponds to the one studied by Jackiw-Rebbi \cite{Jackiw-Rebbi}, and system binds \emph{localized} Majorana fermions, where $m$ changes its sign. \cite{ghaemi-aswin, seradjeh-grosfeld}. Appearance of such Majorana fermions can also be justified in the following way. The components of the s-wave order parameters $\vec{\Delta}_s$ and the Dirac mass ($m$) mutually anticommutes and constitute an $O(3)$ vector, which also anticommute with the non-interacting Dirac Hamiltonian $H_{gen}[0,0]$, when $\mu=0=h_\pm$. Thus together a line vortex of s-wave pairing and domain wall of Dirac mass constitute a {\em hedgehog} in three spatial dimensions, and binds localized Majorana fermions at the end point of line vortex or where $m$ changes its sign\cite{Jackiw-Rebbi}. However, $\vec{\Delta}_T$ and $m$ do not anticommute with each other, and no such Majorana fermion can exist near the domain wall of $m$. If, on the other hand, the system allows a domain wall of parity and time-reversal odd Dirac mass ($m_{PT}$), which together with $\vec{\Delta}_T$, constitute another $O(3)$ vector, localized Majorana fermions can then be realized at the end points of the line vortex of the parity-odd, topological superconductor. 
\\

Our analysis can be applied to an interesting problem of chiral anomaly for the gapless one-dimensional modes along the vortex line of an axial superfluid, which was considered by Callan \& Harvey.\cite{callan-harvey} They have considered the following model
\begin{equation}
H_{ax}=\sum_{j=1}^{3}\gamma_0\gamma_j\left (-i\partial_j-e A_j \right)+ \Delta (r) \gamma_0\exp \left( i \theta \gamma_0 \gamma_5 \right),
\end{equation}
where we have used five mutually anticommuting $\gamma$-matrices, and $A_j$ is the electromagnetic vector potential, and $e$ is the electron's charge. When, we consider a line $n$ vortex of the axial superfluid order parameter along the $z$ direction ($\theta=n\phi$), there are $n$ gapless chiral one-dimensional modes. This number of modes is tied with the Weinberg's $Z$ index theorem for the underlying Jackiw-Rossi problem in the $x-y$ plane. The chiral one-dimensional modes in the presence of the electric field along the $z$ direction, give rise to a nondissipative electric current along the $z$ direction, determined by one-dimensional chiral anomaly $j_z=n \times e^2 E_z/(2\pi)$. This current in turn is supplied radially from the bulk into the vortex core, which is captured by the following axion electrodynamics term:
\begin{equation}
\mathcal{L}_{axion}=-\frac{e^2}{8 \pi^2} \; \int d^4x \epsilon^{\mu \nu \rho \lambda} \partial_{\mu}\theta \; A_{\nu} \; \partial_{\rho}A_{\lambda}.
\end{equation}
Now we may add various fermion bilinears to the above model, which can still support gapless modes along the vortex line. According to the construction of $H_{gen}^{JR}$ in the previous sections, we can add $\gamma_5 \lambda$ and $i\gamma_1 \gamma_2 \chi$, which respectively describe an axial chemical potential, and the third component of the space-like axial vector (these terms break the Lorentz and CPT symmetries). Under this circumstance there is a single gapless mode only for the odd vorticity, if $\Delta_{0}^2+\lambda^2>\chi^2$. The value of the amplitude $\Delta (r)$ at radial infinity has been chosen to be $\Delta_0$. Consequently, we can find nondissipative current only for the odd vorticity, under generic values of these parameters. Accordingly, the bulk axion term, which is usually computed through the Goldstone-Wilczek formula,\cite{goldstone-wilczek} has to be modified to capture the $Z_2$ index.\\

The condition $\Delta_{0}^2+\lambda^2>\chi^2$ has a simple physical meaning in terms of the uniform model's band structure. If $\chi=0$, the Kramer's degeneracy is lifted by $\lambda$, but keeping the spectrum fully gapped. On the other hand, for $\lambda=0$, the spectrum is fully gapped only when $\Delta >\chi$. For $\chi>\Delta$ we have a Weyl semimetal phase, which does not support the vortex zero modes. In the Weyl semimetal phase, there are chiral surface states, which lead to anomalous Hall conductivity and chiral magnetic conductivity.\cite{tewari-goswami} The anomalous transport properties in the gapless phase are also captured by appropriate axion electrodynamics terms, which are not related to vortex zero modes. This is also interesting to note that the number of gapless modes also controls an associated gravitational anomaly formula, and our work suggests its $Z_2$ modification in the presence of $\lambda$ and/or $\chi$. We also note that the axial vector $\chi$ breaks the spatial rotational symmetry, and for this reason the zero modes can only be found for a line vortex aligned with the axial vector.

\acknowledgements
B. R. and P. G. were supported at National High Magnetic Field Laboratory by NSF cooperative agreement No.DMR-0654118, the State of Florida, and the U. S. Department of Energy. Both authors thank the Aspen Center for Physics, where this work was completed during ``2013 Aspen Winter Conference on Topological States of Matter''. B.R. also thanks \'{E}cole de Physique, Les Houches for hospitality during the summer school ``Strongly interacting quantum systems out of equilibrium", where a part of this work was done.

\appendix

\section{Detail of $Z_2$ index theorem of generalized Jackiw-Rossi Hamiltonian}

In this appendix, we present some additional details of the derivation of the $Z_2$ index theorem associated with the existence of the zero energy states in the spectrum of the generalized Jackiw-Rossi Hamiltonian, defined in Eq. (\ref{genJR}).
  
\subsection{Even vorticity}
Upon substituting the ansatz for the zero energy states as in Eq.(\ref{multiangans}) with the constraints over the angular momenta [see Eq. (\ref{evenvortcont})], into Eq. (\ref{eqgenJr}), we obtain the following coupled set of differential equations: 
\begin{eqnarray}
\label{deff1e} \bigg( \partial_r &+& \frac{l}{r} \bigg) f_1(r)+ \Delta_r g_1(r) + (\lambda+\chi) f_2 (r) = 0, \nonumber \\
\label{deff2e}\bigg( \partial_r &-& \frac{l-1}{r} \bigg) f_2(r)+ \Delta_r g_2(r) - (\lambda-\chi) f_1 (r) = 0,\nonumber \\
\label{defg1e}\bigg( \partial_r &-& \frac{2 s -l}{r} \bigg) g_2(r)+ \Delta_r f_2(r) - (\lambda-\chi) g_1 (r) = 0,\nonumber \\
\label{defg2e}\bigg( \partial_r &+& \frac{2 s+1 -l}{r} \bigg) g_1(r)+ \Delta_r f_1(r) + (\lambda+\chi) g_2 (r) = 0, \nonumber\\
\end{eqnarray}  	
where
\begin{equation}\label{evenvortL}
l=0,1, \cdots, (2s-1).
\end{equation}
Therefore, with even vorticity, $n=2 s$, there may exist $n$ number of zero energy states.

\emph{Near origin ($r \to 0$)}: As $r \rightarrow 0$, the pairing order parameter $\Delta_r$ vanishes smoothly, and neglecting the contribution from $\Delta_r$, we arrive at the two sets of coupled differential equations. One of them reads as
\begin{eqnarray}\label{r0f1f2}
\bigg( \partial_r + \frac{l}{r} \bigg) f^<_1(r) + (\lambda+\chi) f^<_2 (r) &=& 0, \nonumber \\
\bigg( \partial_r - \frac{l-1}{r} \bigg) f^<_2(r) - (\lambda-\chi) f^<_1 (r) &=& 0.
\end{eqnarray}
The other set of the coupled differential equations is
\begin{eqnarray}\label{r0g1g2}
\bigg( \partial_r + \frac{2 s+1-l}{r} \bigg) g^<_1(r) + (\lambda+\chi) g^<_2 (r) &=& 0, \nonumber \\
\bigg( \partial_r - \frac{2 s-l}{r} \bigg) g^<_2(r) - (\lambda-\chi) g^<_1 (r) &=& 0.
\end{eqnarray}
Solutions of these differential equations are shown in Eqs.(\ref{r0f1solEmain})-(\ref{f2g2relationr0}).

\emph{Far from origin ($r \to \infty$)}: As $r \rightarrow \infty$, the pairing amplitude $\Delta_r$ approaches the asymptotic value $\Delta_0$, and neglecting all the terms proportional to $1/r$ in Eq.~(\ref{deff1e}), we obtain a new set of four coupled differential equations
\begin{eqnarray}\label{rinfeqEv}
\partial_r f^>_1(r)+ \Delta_0 g^>_1(r)+ (\lambda+\chi)f^>_2(r) &=& 0, \nonumber \\
\partial_r g^>_1(r)+ \Delta_0 f^>_1(r)+ (\lambda+\chi)g^>_2(r) &=& 0,  \nonumber  \\
\partial_r f^>_2(r)+ \Delta_0 g^>_2(r)- (\lambda-\chi)f^>_1(r) &=& 0,  \nonumber \\
\partial_r g^>_2(r)+ \Delta_0 f^>_2(r)- (\lambda-\chi)g^>_1(r) &=& 0.
\end{eqnarray}
Notice that far away from the origin the differential equations are independent of the angular momenta $(l,m,p,q)$. The above four equations reduce to two sets coupled differential equations in terms of new variables, defined as \cite{herbut-lu-ansatz}
\begin{equation}\label{symmasymm}
F_\pm (r)=f^>_1(r) \pm g^>_1(r); \:\: \mbox{and} \:\: G_\pm (r)=f^>_2(r) \pm g^>_2(r).
\end{equation}
In terms of these variables the set of equations in Eq. (\ref{rinfeqEv}) becomes
\begin{eqnarray}\label{combEqEv}
\partial_r F_\pm (r) \pm \Delta_0 F_\pm (r) + (\lambda+ \chi) G_\pm (r) &=& 0, \\
\partial_r G_\pm (r) \pm \Delta_0 G_\pm (r) - (\lambda- \chi) F_\pm (r) &=& 0.
\end{eqnarray}
The solution of these new equations can in general be written as
\begin{eqnarray}\label{gensolcomb}
F_+=\sum_{\sigma=\pm} C_{\sigma} \; \exp \left (\alpha^F_\sigma r \right), \: G_+=\sum_{\sigma=\pm} C'_{\sigma} \; \exp \left (\alpha^F_\sigma r \right), \nonumber \\
F_-=\sum_{\sigma=\pm} \tilde{C}_{\sigma} \; \exp \left (\alpha^G_\sigma r \right), \:  G_-=\sum_{\sigma=\pm} \tilde{C}'_{\sigma} \; \exp \left (\alpha^G_\sigma r \right), \nonumber \\
\end{eqnarray}
where
\begin{equation}\label{rootsrinfEv}
\alpha^F_\sigma=-\Delta_0 + \sigma \sqrt{\chi^2-\lambda^2}; \:
\alpha^G_\sigma=\Delta_0 + \sigma \sqrt{\chi^2-\lambda^2},
\end{equation}
with $\sigma= \pm$. The arbitrary coefficients appearing in the solutions are related according to
\begin{eqnarray}\label{coeff}
C'_{\sigma}= - \sigma \; C_{\sigma} \sqrt{\frac{\chi-\lambda}{\chi+\lambda}}, \:
\tilde{C}'_{\sigma}= \sigma \; \tilde{C}_{\sigma} \sqrt{\frac{\chi-\lambda}{\chi+\lambda}}.
\end{eqnarray}
We are interested only in those solutions which decay exponentially as $r \rightarrow \infty$, so that they are normalizable. Depending on the relative strength of $\Delta_0$, $\lambda$ and $\chi$, the solutions can take different forms.

If $\chi^2<\lambda^2$, and  $(\chi^2-\lambda^2)<\Delta^2_0$, $\alpha^F_\pm <0$, but $\alpha^G_\pm>0$. Therefore, normalizibility of the solutions demands $F_-=G_-=0$. In terms of the original functions, the solutions are given in Eq.~(\ref{fginfCI}). If on the other hand, $\chi^2-\lambda^2 >\Delta^2_0$, $\alpha^{F/G}_- <0$, whereas $\alpha^{F/G}_+ >0$. Hence, for normalizable solutions $C_+=\tilde{C}_+=0$, and the corresponding solutions are given in Eq.~(\ref{fginfCII}).

\subsection{Odd vorticity}
When the vorticity of a point vortex is odd, there exists one possible zero energy state, with single angular momentum, as shown in Eq.~(\ref{singleangans}). Inserting this ansatz into Eq.~(\ref{eqgenJr}), we obtain the following coupled differential equations for $f(r)$ and $g(r)$: 
\begin{eqnarray}\label{fg}
\bigg( \partial_r + \frac{l}{r}\bigg) f(r) + \Delta_r f(r) + (\lambda+\chi)g(r) &=&0, \nonumber \\
\bigg( \partial_r + \frac{1-l}{r}\bigg) g(r) + \Delta_r g(r) - (\lambda-\chi)f(r) &=&0. \nonumber \\
\end{eqnarray}
Following the same strategy, mentioned in the last section, we find that as $r \rightarrow 0$, $f(r)$ is same as $f^<_1(r)$, whereas $g^<(r)$ is given by 
\begin{equation}
g^<(r)=-\frac{1}{\lambda+\chi} \left( \partial_r + \frac{l}{r}\right) f^<(r).
\end{equation}   
At large distances, the above differential equations take the form 
\begin{eqnarray}\label{fgrinfeq}
\partial_r f^>(r) + \Delta_0 f^>(r) + (\lambda+\chi) g^>(r) &=& 0, \nonumber \\
\partial_r g^>(r) + \Delta_0 g^>(r) - (\lambda-\chi) f^>(r) &=& 0,
\end{eqnarray}
which can be solved by using the ansatz $f^>(r)=c \exp{(\alpha r)}$. The solutions of these two equations are presented in Eq.~(\ref{fginfsol}).

The exact solution of the zero energy states from Eq. (\ref{fg}) can be found upon assuming
\begin{equation}\label{exactsolans}
f(r)=\tilde{f}(r) e^{-\int^r_0 \Delta_{r'} dr'}, \: \: \: g(r)=\tilde{g}(r) e^{-\int^r_0 \Delta_{r'} dr'}.
\end{equation}
The functions $\tilde{f}(r)$ and $\tilde{g}(r)$ then satisfy the following equations:
\begin{eqnarray}\label{exactsoleqn}
\bigg( \partial_r + \frac{l}{r} \bigg) \tilde{f}(r) + (\lambda+\chi) \tilde{g}(r) &=& 0, \nonumber \\
\bigg( \partial_r - \frac{l-1}{r} \bigg) \tilde{g}(r) + (\lambda+\chi) \tilde{f}(r) &=& 0,
\end{eqnarray}
yielding
\begin{equation}\label{exactfr}
\tilde{f}(r) = \left\{
\begin{array}{rl}
C_l \; J_{\sqrt{l}} \big[ r\sqrt{\lambda^2-\chi^2} \big] & \text{if} \: \: \lambda > \chi, \\
C_l \; I_{\sqrt{l}} \big[ r\sqrt{\chi^2-\lambda^2} \big] & \text{if} \: \: \lambda < \chi,
\end{array} \right.
\end{equation}
and
\begin{equation}\label{exactgr}
\tilde{g}(r)=-\bigg(\frac{1}{\lambda+\chi}\bigg) \: \bigg( \partial_r + \frac{l}{r} \bigg) \tilde{f}(r).
\end{equation}
Normalizibility of these solutions, depending on the mutual strength of $\lambda$, $\chi$, and $\Delta_0$, has been discussed in the Sec. IIC.

Besides the zero mode with ansatz in Eq.~(\ref{singleangans}), there are additional $2 s$ possible ansatz similar to that in Eq. (\ref{multiangans}). The angular momenta satisfy the constraints in Eq. (\ref{multangcontodd}). Upon imposing those constraints over the angular momenta, one set of coupled differential equations involving the functions $f^<_1(r)$ and $f^<_2(r)$, in the vicinity of the origin assumes the identical form as in Eq. (\ref{r0f1f2}). The other set of coupled differential equations is
\begin{eqnarray}\label{g1g2Evr0eq}
\bigg( \partial_r + \frac{2s +2 -l}{r}\bigg) g^<_1(r) + (\lambda+\chi)g^<_2(r) &=& 0, \nonumber \\
\bigg( \partial_r + \frac{l- 2s -1}{r}\bigg) g^<_2(r) - (\lambda-\chi)g^<_1(r) &=& 0.
\end{eqnarray}
The solutions for $g^<_1(r)$ and $g^<_2(r)$ are shown in Eq. (\ref{g1oddr0sol}). Far from the origin, the differential equations are independent of angular momenta, and takes the form of Eqs.~(\ref{fginfCI}) or (\ref{fginfCII}), depending on the mutual strength of $\Delta_0$, $\chi$, and $\lambda$.

\section{Zero-energy states at $\lambda=\pm \chi$}

We here present the detail solutions of the zero energy states with an underlying antivortex for $\lambda=\pm \chi$. When $\lambda=-\chi$, one can write the equation for the zero energy modes for even-vorticity ($n=2s$) from Eq. (\ref{deff1e}). Written slightly differently, they read as
\begin{eqnarray}
&& \bigg( \partial_r + \frac{p+1}{r} \bigg) f_1(r)+ \Delta_r g_1(r)=0, \nonumber \\
&& \bigg( \partial_r - \frac{p}{r} \bigg) f_2(r) + \Delta_r g_2 (r)- 2 \lambda f_1 (r)=0, \nonumber \\
&& \bigg( \partial_r - \frac{2s-1-p}{r} \bigg) g_2 (r)+ \Delta_r f_2(r)-2 \lambda g_1(r)=0, \nonumber \\
&& \bigg( \partial_r + \frac{2s-p}{r} \bigg) g_1(r) + \Delta_r f_1(r)=0,
\end{eqnarray}
where $p$ is a positive definite integers, and takes the values
\begin{equation}\label{pvalues}
p=0,1, \cdots, 2s-1.
\end{equation}
In the vicinity of the origin, where $\Delta_r \rightarrow 0$, $f^{<}_1(r) =\tilde{c}_1 r^{-(p+1)}$, and $g^{<}_1(r)=\tilde{c}_2 r^{-(2s-p)}$. Hence, normalizable zero energy modes can only be found when $\tilde{c}_1=\tilde{c}_2=0$. On the other hand, in the vicinity of the origin,
\begin{equation}
f^{<}_2(r)= c_1 r^{p}, \: \: g^{<}_2(r)=c_2 r^{2s-1-p},
\end{equation}
are well behaved functions for all $p$ given in Eq. (\ref{pvalues}). Far away from the origin, these two functions are $f^{>}_2(r)=g^{>}_2(r)=c \exp{(-\Delta_0 r)}$. Two out of three arbitrary constants $(c_1,c_2,c)$ can be fixed by imposing the boundary conditions
\begin{equation}\label{lambdaxiboundary}
f^{<}_1(r=\xi)=f^{>}_1(r=\xi), \: \: g^{<}_1(r=\xi)=g^{>}_1(r=\xi),
\end{equation}
while the remaining one is determined by the overall normalization factor. Hence there are $n=2s$ number of zero energy states with underlying antivortex, when $\lambda=-\chi$. In terms of the original functions in Eq. (\ref{multiangans}), $V(\vec{x}) =0$ for the zero energy modes. Therefore, the zero energy modes are eigenstate of the operator $\Gamma_5=\sigma_0 \otimes \sigma_3$ with eigenvalue $+1$. These solutions match exactly with the ones found by Jackiw and Rossi, in the absence of gauge fields \cite{JR-original}.

If, on the other hand, $\lambda=\chi$, we have
\begin{eqnarray}
f^{<}_2(r)&=&\tilde{c}_2 r^{p}, \; f^{<}_1(r)=-\frac{\tilde{c}_2 \lambda}{p+1} r^{p+1}, \nonumber \\
g^{<}_2(r)&=& c'_2 r^{2s-1-p},  \; g^{<}_1(r)=\frac{c'_2 \lambda}{p-2s} r^{2s-p},
\end{eqnarray}
where $\tilde{c}_2, c'_2$ are arbitrary constants. Far away from the origin, the functions behave as
\begin{eqnarray}
f^{>}_2(r)=g^{>}_2(r)&=& c \; \exp{(-\Delta_0 r)}, \nonumber \\
f^{>}_1(r)=g^{>}_1(r)&=& - 2 c \lambda r \; \exp{(-\Delta_0 r)},
\end{eqnarray}
where $c$ is also an arbitrary constant. After satisfying the boundary conditions, for example the one shown in Eq. (\ref{lambdaxiboundary}), two out of three arbitrary constants, $(\tilde{c}_2,c'_2, c)$ are fixed. With only one arbitrary constant, it is now impossible to satisfy two similar boundary conditions for $f_2(r)$ and $g_2(r)$. Therefore, when $\lambda=\chi$, there is no zero energy state when the underlying antivortex has even vorticity.

Let us now consider the antivortex with odd vorticity, $n=2s +1$. We first focus on $2 s$ ansatz of the form Eq. (\ref{multiangans}). The equations of the zero energy modes then read as  
\begin{eqnarray}
&& \bigg( \partial_r + \frac{p+1}{r} \bigg) f_1(r)+ \Delta_r g_1(r)=0, \nonumber \\
&& \bigg( \partial_r - \frac{p}{r} \bigg) f_2(r) + \Delta_r g_2 (r)- 2 \lambda f_1 (r)=0, \nonumber \\
&& \bigg( \partial_r - \frac{2s-p}{r} \bigg) g_2 (r)+ \Delta_r f_2(r)-2 \lambda g_1(r)=0, \nonumber \\
&& \bigg( \partial_r + \frac{2s-p+1}{r} \bigg) g_1(r) + \Delta_r f_1(r)=0,
\end{eqnarray}
when $\lambda=-\chi$. For the normalizable zero energy modes, we find $f_1(r)=g_1(r)=0$. The remaining two functions in the vicinity of the origin are given by
\begin{equation}
f^{<}_2(r)=c_1 r^{p}, \; g^{<}_2(r)=c_2 r^{2s-p},
\end{equation}
and far away from origin they are
\begin{equation}
f^{>}_2(r)=g^{>}_2(r)=c \exp{(-\Delta_0 r)}.
\end{equation}
Hence, for $\lambda=-\chi$, there exists $2 s$ number of zero energy states with an underlying antivortex of vorticity $2 s+1$ of the form Eq. (\ref{multiangans}). These solutions are also identical to the ones one find when $\lambda=\chi=0$ \cite{JR-original}. However, when $\lambda=\chi$, there is no zero energy mode of the form Eq. (\ref{multiangans}).

Next we solve the zero energy mode with the ansatz of the form Eq. (\ref{singleangans}), when $\lambda=\pm \chi$. One can also write the Eq. (\ref{fg}) for the zero energy states as
\begin{eqnarray}
&& \bigg( \partial_r + \frac{p+1}{r}\bigg) f(r) + \Delta_r f(r) + (\lambda+\chi)g(r) =0, \nonumber \\
&& \bigg( \partial_r - \frac{p}{r}\bigg) g(r) + \Delta_r g(r) - (\lambda-\chi)f(r) =0,
\end{eqnarray}
where $p=(n-1)/2$, and $n$(odd) is the vorticity. For $\lambda=-\chi$, the radial functions of the zero modes are given by
\begin{equation}
f(r)=0, \: g(r)= c \; r^{p} \; \exp{\bigg(-\int^r_0 \Delta_{r'}  \; dr' \bigg)},
\end{equation}
where $c$ is an arbitrary constant, similar to the one found in original work by Jackiw-Rossi \cite{JR-original}. With this particular ansatz, we also find normalizable zero energy modes even when $\lambda=\chi$. The radial functions then go as
\begin{eqnarray}
f(r)&=& - c \; \bigg( \frac{\lambda}{p+1} \bigg) \; r^{p+1} \; \exp{\bigg(-\int^r_0 \Delta_{r'}  \; dr' \bigg)}, \nonumber \\
g(r)&=& c \; \exp{\bigg(-\int^r_0 \Delta_{r'}  \; dr' \bigg)}.
\end{eqnarray}
Therefore, the zero energy mode of the form Eq. (\ref{singleangans}), exists whether $\lambda=\chi$ or $\lambda=-\chi$. For $n=1$ or $p=0$, this solution matches exactly with the one shown in Ref. \onlinecite{herbut-lu-genJR}.

\section{Detail of dispersive and vortex zero modes}

\subsection{Topological pairing}

Let us first present some details for the solutions of the gapless dispersive modes and fermionic vortex zero energy state with underlying topological pairing when $\mu=h_+=0$. The coupled differential equations for gapless modes along the vortex core (chosen in $z$ direction) read as
\begin{eqnarray}\label{ELMMT}
(m &+& h_-) \Lambda^+_\uparrow + (-i) e^{-i \phi}\left( \partial_r -\frac{i}{r} \partial_\phi + a_\phi \right) \Lambda^-_\downarrow   \nonumber \\
&+& k_z \Lambda^-_\uparrow + \Delta_r e^{-i \phi} \bigg(\Lambda^-_\downarrow \bigg)^\dagger = E \Lambda^+_\uparrow,  \\
-(m &-& h_-) \Lambda^-_\downarrow + (-i) e^{i \phi}\left( \partial_r + \frac{i}{r} \partial_\phi - a_\phi \right) \Lambda^+_\uparrow \nonumber \\
&-& k_z \Lambda^+_\downarrow- \Delta_r e^{-i \phi} \bigg(\Lambda^+_\uparrow \bigg)^\dagger = E \Lambda^-_\downarrow, \\
(m &-& h_-) \Lambda^+_\downarrow + (-i) e^{i \phi}\left( \partial_r + \frac{i}{r} \partial_\phi - a_\phi \right) \Lambda^-_\uparrow  \nonumber \\
&-& k_z \Lambda^-_\downarrow - \Delta_r e^{-i \phi} \bigg(\Lambda^-_\uparrow \bigg)^\dagger = E \Lambda^+_\downarrow, \\
-(m &+& h_-) \Lambda^-_\uparrow + (-i) e^{-i \phi}\left( \partial_r -\frac{i}{r} \partial_\phi + a_\phi \right) \Lambda^+_\downarrow \nonumber \\
&+& k_z \Lambda^+_\uparrow + \Delta_r e^{-i \phi} \bigg(\Lambda^+_\downarrow \bigg)^\dagger = E \Lambda^-_\uparrow.
\end{eqnarray}
Remaining four equations are simply the complex conjugations of the above four. For $E=k_z$, the various components of the Nambu-Dirac spinor are related according to 
\begin{equation}\label{constEp}
\Lambda^-_\uparrow =\Lambda^+_\uparrow, \Lambda^-_\downarrow = - \Lambda^+_\downarrow,
(\Lambda^-_\uparrow)^\dagger =(\Lambda^+_\uparrow)^\dagger, (\Lambda^-_\downarrow)^\dagger = -( \Lambda^+_\downarrow)^\dagger.
\end{equation}
Upon imposing these constraints, the above four equations reduce to only two, and they take the form
\begin{eqnarray}\label{EPE1T}
(-i) e^{-i \phi}\bigg( \partial_r -\frac{i}{r} \partial_\phi &+& a_\phi \bigg) \Lambda^-_\downarrow + (m+h_-) \Lambda^+_\uparrow  \nonumber \\
&+& \Delta_r e^{-i \phi} \bigg(\Lambda^-_\downarrow \bigg)^\dagger = 0,
\end{eqnarray}
\begin{eqnarray}\label{EPE2T}
(-i) e^{i \phi}\bigg( \partial_r + \frac{i}{r} \partial_\phi &-& a_\phi \bigg) \Lambda^+_\uparrow - (m-h_-) \Lambda^-_\downarrow \nonumber \\
&-& \Delta_r e^{-i \phi} \bigg(\Lambda^+_\uparrow \bigg)^\dagger = 0.
\end{eqnarray}
These two coupled equations can be solved using the ansatz
\begin{equation}\label{solLPPMM}
\lambda^+_\uparrow = {\cal R}(r,0) \; e^{-i \phi} \; g(r), \: \lambda^-_\downarrow = {\cal R}^*(r,0) \; f(r),
\end{equation}
yielding the solution shown in Eq.(\ref{Epktopo}) for $| E=+k_z\rangle$. On the other hand, when we wish to solve the dispersive mode with $E=-k_z$, the components of Nambu spinor are related according to
\begin{equation}\label{Emkcont}
\Lambda^-_\uparrow = - \Lambda^+_\uparrow, \Lambda^-_\downarrow =  \Lambda^+_\downarrow,
(\Lambda^-_\uparrow)^\dagger =- (\Lambda^+_\uparrow )^\dagger, (\Lambda^-_\downarrow )^\dagger = ( \Lambda^+_\downarrow )^\dagger.
\end{equation}
Following the same steps above, we find the other dispersive mode $| E=-k_z \rangle$, also shown in Eq.(\ref{Epktopo}).

Next we present the solution of the fermionic zero modes with an underlying point vortex, when $\mu=h_+=0$. With this particular choice of parameters, after setting $E=k_z=0$, one set of coupled differential equations for the zero mode is obtained from Eqs.~(C1) and (C2), yielding the solution $| \Psi^0_1 \rangle$ shown in Eq.~(\ref{vorttoposol1}). The other set of coupled differential equations for the fermionic zero mode arises from Eqs.~(C3) and (C4), giving the solution $| \Psi^0_2 \rangle$ in Eq.~(\ref{vorttoposol1}).

\subsection{s-wave pairing}
Let us now consider a line vortex along the $z$ direction with underlying s-wave paring. The dispersive mode can be solved analytically when we set $m=\mu=h_\pm=0$. Furthermore, we also turn off the orbital contribution of the gauge field, i.e., $a_\phi=0$. Then the coupled differential equations read as
\begin{eqnarray}\label{swaveeqn}
(-i) \left( \partial_r - \frac{i}{r} \partial_\phi \right) \Lambda^-_\downarrow + k_z \Lambda^-_\uparrow + e^{-i \phi} \Delta_r \left( \Lambda^+_\downarrow \right)^\dagger= E \Lambda^+_\uparrow, \nonumber \\
(-i) \left( \partial_r + \frac{i}{r} \partial_\phi \right) \Lambda^-_\uparrow - k_z \Lambda^-_\downarrow + e^{-i \phi} \Delta_r \left( \Lambda^+_\uparrow \right)^\dagger= E \Lambda^+_\downarrow, \nonumber \\
(-i) \left( \partial_r - \frac{i}{r} \partial_\phi \right) \Lambda^+_\downarrow + k_z \Lambda^+_\uparrow + e^{-i \phi} \Delta_r \left( \Lambda^-_\downarrow \right)^\dagger= E \Lambda^-_\uparrow, \nonumber \\
(-i) \left( \partial_r + \frac{i}{r} \partial_\phi \right) \Lambda^+_\uparrow - k_z \Lambda^+_\downarrow + e^{-i \phi} \Delta_r \left( \Lambda^-_\uparrow \right)^\dagger= E \Lambda^-_\downarrow. \nonumber \\
\end{eqnarray}
The remaining four equations are the Hermitian conjugates of above four. Upon imposing the constraints over the spinor components, as shown in Eq. (\ref{constEp}), the above four equation reduces to \emph{two} decoupled equations. For $E=+k_z$, they read as
\begin{eqnarray}
i e^{-i \phi} \left( \partial_r - \frac{i}{r} \partial_\phi \right) \Lambda^-_\downarrow + e^{-i \phi} \Delta_r \left( \Lambda^-_\downarrow \right)^\dagger &=& 0, \\
i e^{i \phi} \left( \partial_r + \frac{i}{r} \partial_\phi \right) \Lambda^+_\uparrow + e^{-i \phi} \Delta_r \left( \Lambda^+_\uparrow \right)^\dagger &=& 0.
\end{eqnarray}
These two equations, respectively, yield
\begin{equation}\label{CPM}
\Lambda^-_\downarrow = C^+_- e^{-i \frac{\pi}{4}} e^{-\int^r_0 \Delta_{r'} dr'},
\Lambda^+_\uparrow = \frac{C^+_+}{r} e^{-i \frac{\pi}{4}} e^{-\int^r_0 \Delta_{r'} dr'},
\end{equation}
where $C^+_\pm$ are the arbitrary constants. However, to keep the second solution well behaved near the origin, we have to set $C^+_+=0$, therefore $\Lambda^+_\uparrow \equiv0$. On the other hand, the constraints in Eq. (\ref{Emkcont}) yield the following two decoupled equations for the dispersive mode with $E=-k_z$:
\begin{eqnarray}
(-i) e^{-i \phi} \left( \partial_r - \frac{i}{r} \partial_\phi \right) \Lambda^-_\downarrow + e^{-i \phi} \Delta_r \left( \Lambda^-_\downarrow \right)^\dagger &=& 0, \nonumber \\ \\
(-i) e^{i \phi} \left( \partial_r + \frac{i}{r} \partial_\phi \right) \Lambda^+_\uparrow + e^{-i \phi} \Delta_r \left( \Lambda^+_\uparrow \right)^\dagger &=& 0,\nonumber \\
\end{eqnarray}
giving
\begin{equation}\label{CMM}
\Lambda^-_\downarrow = (i) \; C^-_- e^{-i \frac{\pi}{4}} e^{-\int^r_0 \Delta_{r'} dr'},
\Lambda^+_\uparrow = \frac{C^-_+}{r} e^{i \frac{\pi}{4}} e^{-\int^r_0 \Delta_{r'} dr'},
\end{equation}
respectively, where $C^-_\pm$ are also arbitrary constants. For the solutions to be well behaved in the vicinity of the origin, we need to set $C^-_+=0$, and thus $\Lambda^+_\uparrow=0$ once again. In conjunction with these constraints, the above solutions in Eqs. (\ref{CPM}) and (\ref{CMM}), yield two dispersive modes with underlying s-wave order, shown in Eq.~(\ref{swave1D}).


\begin{thebibliography}{99}
\bibitem{explanationclifford} The Clifford algebra of $4=2^q$-dimensional mutually anticommuting matrices is closed by $2 q+1=5$ matrices. Out of them, $q+1=3$ can can chosen to be real, while $q$ to be imaginary. See also Refs. \onlinecite{okubo, herbut-lu-genJR}.
\end{thebibliography}
\end{document}